\documentclass[journal, onecolumn, 12pt]{IEEEtran}
\usepackage[centertags]{amsmath}
\usepackage{amsfonts}
\usepackage{amssymb}
\usepackage{amsthm}
\usepackage{courier}
\usepackage{newlfont}
\usepackage{graphicx,subfigure}
\usepackage{array,multirow}
\newcommand{\ber}   {\begin{eqnarray}}
\newcommand{\eer}   {\end{eqnarray}}
\newcommand{\beq}   {\begin{equation}}
\newcommand{\eeq}   {\end{equation}}
\newcommand{\bmx}       {\begin{bmatrix}}
\newcommand{\emx}       {\end{bmatrix}}

\newcommand{\calC} {\cal{C}}
\newcommand{\calA} {\cal{A}}
\newcommand{\calE} {\cal{E}}

\newtheorem{theorem}{Theorem}
\newtheorem{lemma}{Lemma}
\newtheorem{corollary}{Corollary}
\newtheorem{definition}{Definition}

\newtheorem{remark}{Remark}

\begin{document}

\title{An Upper Bound on Relaying over Capacity Based on Channel Simulation}

\author{\IEEEauthorblockN{Feng Xue} \\
\IEEEauthorblockA{Intel Labs, 2200 Mission College Blvd, Santa Clara, CA 95054
\\
Email: fengxue@ieee.org}
\\
September 2012
}

\maketitle

\begin{abstract}
 The upper bound on the capacity of a 3-node discrete memoryless relay channel is considered, where a source $X$ wants to send information to destination $Y$ with the help of a relay $Z$. $Y$ and $Z$ are independent given $X$, and the link from $Z$ to $Y$ is lossless with rate $R_0$. A new inequality is introduced to upper-bound the capacity when the encoding rate is beyond the capacities of both individual links $XY$ and $XZ$. It is based on generalization of the blowing-up lemma, linking conditional entropy to decoding error, and channel simulation, to the case with side information. The achieved upper-bound is strictly better than the well-known cut-set bound in several cases when the latter is $C_{XY}+R_0$, with $C_{XY}$ being the channel capacity between $X$ and $Y$. One particular case is when the channel is statistically degraded, i.e., either $Y$ is a statistically degraded version of $Z$ with respect to $X$, or $Z$ is a statistically degraded version of $Y$ with respect to $X$. Moreover in this case, the bound is shown to be explicitly computable. The binary erasure channel is analyzed in detail and evaluated numerically.
 \end{abstract}


\section{Introduction}
The relay channel model was first formulated by Van-der Meulen \cite{Meulen1971} in 1971, consisting a source $X$, a relay $Z$, and a destination $Y$. The relay transmits a signal $X_1$ based on its observation to help $Y$. As a basic building block of general communication networks, it has since then attracted much research interests; see e.g. \cite{ElGamal2010isit} and references therein.

A set of achievability results were introduced by Cover and El Gamal \cite{CoverElGamal1979}. Decode-forward and compress-forward are two basic achievability methods.
Several capacity results were established for degraded, reverse degraded \cite{CoverElGamal1979}, semi-deterministic \cite{ElGamalAref1982}, and deterministic \cite{Kim2008}  channels.  They are all based on achieving the well-known cut-set bound with certain coding scheme; see e.g. Chapter 14 of \cite{CoverThomas1991}.

In general, however, the cut-set bound seems not tight. A result on this was shown by Zhang in 1988 \cite{Zhang1988} for the channel depicted in Figure 1. The link from the relay to the destination is assumed to be lossless with fixed rate $R_0$. $Y$ and $Z$ are conditionally independent given $X$. Furthermore, $Y$ is a statistically degraded version of $Z$ with respect to $X$. In other words, $X$-$Z$-$Y$ can be re-described as a Markov chain.
By applying the blowing up lemma \cite{AlswedeGacsKorner1976}, it is shown by contradiction that the cut-set bound cannot be tight.  However, it is still unknown how loose the bound is. In \cite{AleksicRazaghiYu2009}, a specific  class of modulo additive noise relay channels is considered. The relay observes a noisy version of the noise corrupting the signal at the destination. The capacity is established and shown to be strictly lower than the cut-set bound. To the best knowledge of the author, there is no general upper-bound tighter than the cut-set bound for the relay channel.

%
\begin{figure}[htbp]
  \centering
    \includegraphics[height=4cm]{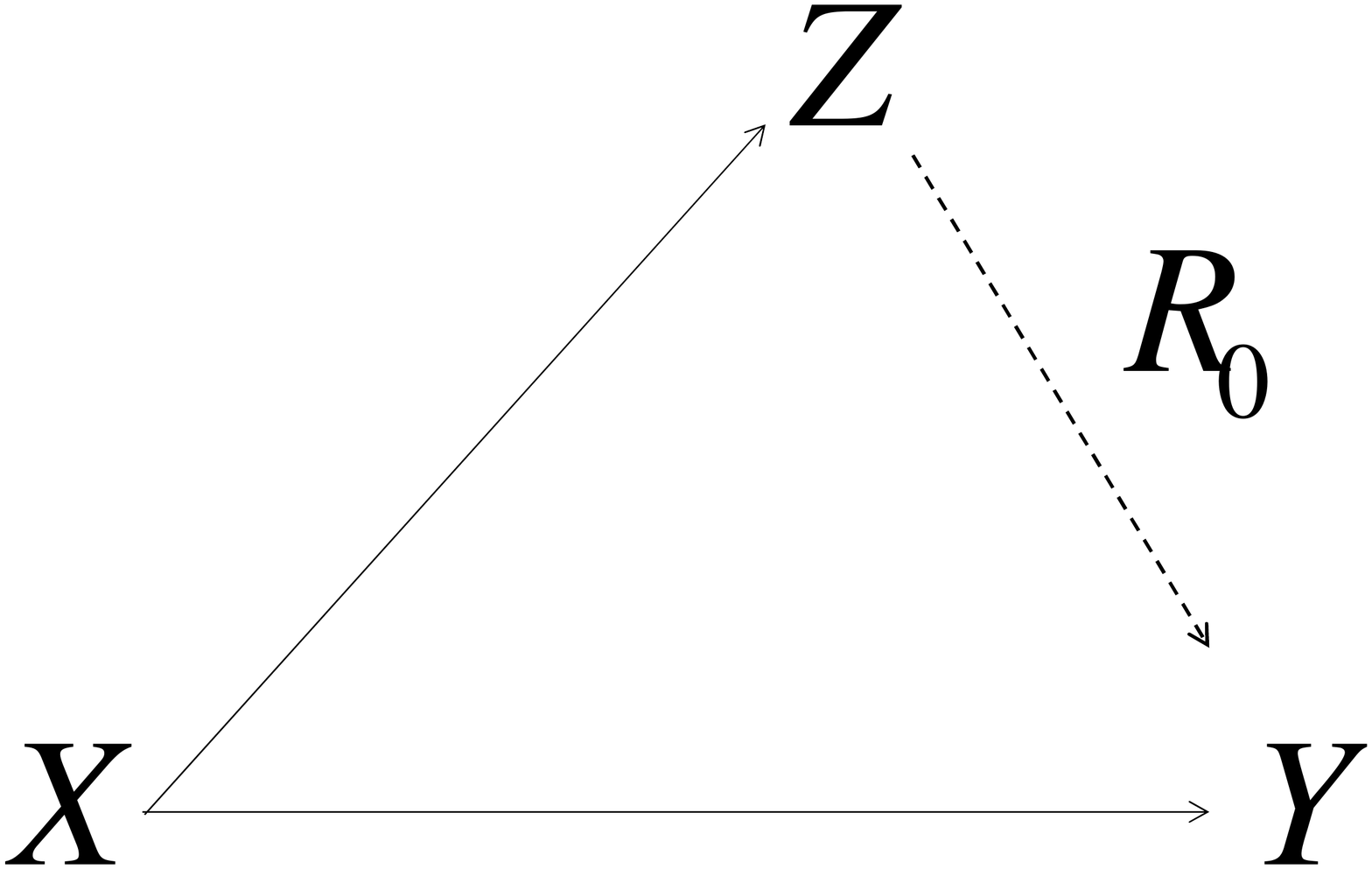}
    \caption{Relay Network with lossless relay-destination link}
    \label{Fig:system0}
\end{figure}
\par

In this paper, we consider improving the cut-set bound for the channel depicted in Figure 1, similar to \cite{Zhang1988}. Nodes $Y$ and $Z$ are independent given $X$, and the link from the relay $Z$ to the destination $Y$ is lossless with rate $R_0$. Specifically, in a transmission of $n$ channel uses, a ``color" in $\{1, 2, \cdots, 2^{nR_0} \}$ can be sent to $Y$ without error.

The cut-set bound for this relay channel is
\begin{eqnarray}
\max_{p(x)} \min \{I(X; Y)+R_0, I(X; Y, Z) \}.
\end{eqnarray}
It equals $C_{XY}+R_0$ in many cases when $R_0$ is small, where $C_{XY}$ denotes the channel capacity between $X$ and $Y$. This is based on the following observation. Suppose under input distribution $p^*(x)$, $I(X; Y)$ becomes $C_{XY}$.  Then as long as $I(X; Y, Z)>I(X; Y)$ under $p^*(x)$, the cut-set bound is $C_{XY}+R_0$ whenever $R_0$ is such that $C_{XY}+R_0< I(X; Y, Z)$.

In this paper, a new bounding technique is introduced. It leads to an explicit and strictly-better upper bound than $C_{XY}+R_0$ when $R_0>0$ and the encoding rate is beyond both $C_{XY}$ and $C_{XZ}$. We present the following results specifically.
First, we show an explicitly computable bound for the case when the channel is statistically degraded. That is, $Z$ is a statistically degraded version of $Y$ with respect to $X$, or $Y$ is a statistically degraded version of $Z$. The bound is strictly lower than $C_{XY}+R_0$, and thereby improves the result in \cite{Zhang1988} directly. As an example, the binary erasure channel is analyzed in detail.
Secondly, by extending the method of channel simulation\cite{HanVerdu1993, Cuff2008}, we generalize the results to cases when the channel is not necessarily degraded.

The essential idea of this bounding technique is to introduce a fundamentally new inequality on any feasible rate, in addition to Fano's inequality \cite{CoverThomas1991}. In our case, Fano's inequality manifests as $R<C_{XY}+R_0-\frac{1}{n} H(\hat{Z}^n|X^n)$, where $\hat{Z}^n$ denotes the color the relay sends to $Y$ and $X^n$ is the codeword.
Our new inequality is established by combining two observations for any feasible code:
\begin{itemize}
\item
First, it is known that the decoding probability for a memoryless channel decays exponentially when the encoding rate is beyond capacity. This is universal and independent of the encoding/decoding technique. Moreover, the exponent is explicitly computable \cite{Arimoto1973}.
\item
Secondly, any feasible rate and associated encoding/decoding scheme provide a way for node\footnote{We will also consider the case where node $Z$ makes the guess.} $Y$ to guess the codeword $X^n$ based {\em solely} on its own signal $Y^n$, as follows. Since the rate is feasible, there must be a decoding function which maps $Y^n$ and the color $\hat{Z}^n$
to the correct codeword. So node $Y$ only needs to guess the color $\hat{Z}^n$. To accomplish this, one notices that,
when $\frac{1}{n} H(\hat{Z}^n|X^n)$ is close to zero, the color $\hat{Z}^n$ turns to be a deterministic function of $X^n$, even though the $XZ$ channel is random.
So if $Y$ can generate a random variable $\tilde{Z}^n$ with the same distribution of $Z^n$ given $X^n$,
there is a good chance for $Y$ to guess the ``color".  This guessing  is achieved by generalizing the blowing-up lemma. Overall, the probability of successful decoding can be determined.
\end{itemize}
Based on the first observation, the probability of success in the second observation must be less than the universal bound, and thereby the second inequality establishes. With Fano's and our inequalities at hands, it will be clear that the second inequality becomes active when $\frac{1}{n} H(\hat{Z}^n|X^n)$ is small,  and it bounds the rate away from the cut-set bound.

One critical step in our method is to generate a random variable with the same distribution of $Z^n$ (or $Y^n$) given $X^n$.
When the channel is statistically degraded, this task is straightforward.
In more general cases, method based on {\em channel simulation} \cite{HanVerdu1993, Cuff2008} can be applied.
Channel simulation \cite{HanVerdu1993, Cuff2008} aims generating random variable in an ``efficient" way. In our case, where a side information is available (e.g. $Y^n$ when $Y$ needs to ``simulate" $Z^n$), a generalization of the known results is derived and applied for achieving the new inequality.

The rest of the paper is organized as follows. Section II introduces the basic definitions, notations and a well-known bound on decoding probability when encoding rate is beyond channel capacity. Section III generalizes the blowing up lemma and links to conditional entropy. Section IV applies it to characterize the bound for the case when $Y$ and $Z$ are i.i.d. given $X$; it also takes the binary erasure channel for detailed derivation. Section V subsequently generalizes the results to the case when the channel is statistically degraded. Section VI presents the channel simulation and generalizes it to the cases when side information is available. This is later applied to our relay channel in Section VII to achieve a general bound. Finally, Section VIII concludes with some remarks.

\section{Definitions, notations and a well-known bound on decoding probability}
The memoryless relay channel we consider consists of three nodes, sender $X$, relay $Z$ and destination $Y$, defined by the conditional distribution $p(y, z |x)$. $Y$ and $Z$ are independent given $X$, i.e., $p(y, z|x)=p(y|x) p(z|x)$. The values of $X$, $Y$ and $Z$ are from finite spaces $\Omega_X$, $\Omega_Y$ and $\Omega_Z$ respectively. Correspondingly, for a transmission of length $n$, the code word $x^n$ is chosen from $\Omega_X^n$, the product space of $\Omega_X$, and the received observations are $y^n \in \Omega_Y^n$ and $z^n \in \Omega_Z^n$, respectively. The link from the relay to the destination is a lossless link with rate $R_0$. Namely, for a transmission of $n$ channel uses, a number from $\{1, 2, \cdots, 2^{nR_0}\}$ can be sent to $Y$ without error.

A {\em coding strategy of rate $R$} for $n$ channel uses is defined by a 3-tuple $(\calC^{(n)}, g_n(z^n), f_n(\hat{z^n}, y^n))$.
Set $\calC^{(n)}:=\{x^n(m), m=1, \cdots, 2^{nR} \}$ is the code book at the source $X$. Node $X$ chooses one codeword uniformly from the set and transmits to the channel.
Function $g_n(z^n)$ is the encoding function at the relay $Z$, which is a function mapping an observation $z^n$ to $\hat{z^n}$, which is a ``color" $j$ in $\{1, 2, \cdots, 2^{nR_0}\}$. In this paper, we use $\hat{z^n}$ to denote this mapping function, and call the set $\{1, 2, \cdots, 2^{nR_0}\}$ the {\em color set}.
Function $f_n(\hat{z^n}, y^n)$ is the decoding function at the destination $Y$, mapping the color from the relay and the observation $y^n$ to a code word in $\calC^{(n)}$.
All $\calC^{(n)}, g_n(\cdot)$ and $f_n(\cdot)$ are well-known at all nodes.

\begin{definition}
{\em Rate $R$ is feasible} if there exists a sequence of coding strategies of rate $R$, \\
$\{ (\calC^{(n)}, g_n(z^n), f_n(\hat{z^n}, y^n) ), n \geq 1\}$,
such that the successful decoding probability approaches one as $n$ goes to infinity. That is,
$$\lim_n Pr(f_n(\hat{Z^n}, Y^n) = X^n) = 1.$$
\end{definition}

We introduce several notations here.
\begin{itemize}
\item
$C_{XY}$ and $C_{XZ}$ are the channel capacities from the channels $X$-$Y$ and  $X$-$Z$, respectively.
\item
The notation $d_H(x_1^n, x_2^n)$ denotes the Hamming distance of two points.
\item
Throughout the paper, $\log$ is with base 2. Also, we reserve the use of the hat symbol $\hat{\omega}$ on top of a random variable solely for the coloring.
\end{itemize}

 We now quote the result on decoding probability when transmitting at rate above a channel's capacity.

\subsection{Decoding Probability When Based on $Y^n$ Only}
Consider only the transmission between $X$ and $Y$, and ignore $Z$. That is, the destination $Y$ wants to decode the codeword by using $Y^n$ only. When the code book has rate above the capacity, it is well-known that the decoding probability approaches zero exponentially fast. The following is shown in  \cite{Arimoto1973}.

\begin{theorem} \label{Arimoto1973} Suppose that a discrete memoryless channel with an input alphabet of K letters $\{a_1, \cdots, a_K\}$ and an output alphabet of J letters $\{b_1, \cdots, b_J\}$ is described by transition probabilities $P_{jk}=p(b_j | a_k)$. Then, for any block length $n$ and any code book of size $M=2^{nR}$, the probability of decoding satisfies
\begin{eqnarray}
\label{arimotoExponent}
Pr(Decoding) \leq 2^{ -n (-\rho R + \min_p \Phi_0(\rho, p))  }, \quad \forall \rho \in [-1, 0),
\end{eqnarray}
where $p$ represents a distribution over the input alphabet $\{ p_k \}$, and
$$\Phi_0(\rho, p):= -\log \left[ \sum_{j=1}^J \left\{ \sum_{k=1}^K p_k P_{jk}^{1/(1+\rho)} \right\}^{(1+\rho)} \right].$$
\end{theorem}

In the paper, we denote the largest exponent as
\begin{eqnarray}
\label{def_decoErrorProb}
\calE(R):=\max_{\rho \in [-1, \,\, 0)} (-\rho R + \min_p \Phi_0(\rho, p))
\end{eqnarray}

\begin{remark}{\em  (\cite{Arimoto1973})}
\label{remarkOnerrexponent}
It is easy to show that $\calE(R) >0$ for any given $R>C_{XY}$. Also note that
$$\lim_{\rho \to 0^-} \frac{1}{\rho} \min_{p} \Phi_0(\rho, p) = \lim_{\rho \to 0^+} \frac{1}{\rho} \min_{p} \Phi_0(\rho, p)=C_{XY}.$$
\end{remark}

\section{Generalizing the Blowing-Up Lemma}

The well-known blowing-up lemma \cite{AlswedeGacsKorner1976, Marton1986} states that if an event ${\calA}^{(n)}$ in a product probability space $\Omega^n$ has probability diminishing slower than exponential, then the event consisting all points that are within a small Hamming distance of ${\calA}^{(n)}$ will have a probability going to one. More precisely, it is the following.
\vspace{2mm}
\begin{lemma}
(The Blowing up Lemma) Let $Q_1, Q_2, \cdots, Q_n$ be independent random variables in a finite space $\Omega$, with distribution $P_{Q_i}$ respectively. Denote random vector $Q^n:=(Q_1, \cdots, Q_n)$ and the joint distribution $P_{Q^n}:=\Pi_{i=1}^n P_{Q_i}$. Suppose there exist $\epsilon_n \to 0$ and event ${\calA}^{(n)} \in \Omega^n$ such that $Pr(Q^n \in {\calA}^{(n)}) \geq 2^{-n \epsilon_n}$. Then there exist  $\delta_n, \eta_n$ going to 0 such that $Pr(Q^n \in \Gamma_{n \delta_n} ({\calA}^{(n)})) \geq 1-\eta_n$, where
 $\Gamma_l({\calA}^{(n)}):=\{x^n: \min_{y^n \in {\calA}^{(n)}} d_H(x^n, y^n) \leq l\}$ is the ``blown-up" set.
\end{lemma}

This lemma can be generalized to the case without requirement on the event probability as follows.
\vspace{2mm}
\begin{lemma}
\label{BlowupLemma}
Let $Q_1, Q_2, \cdots, Q_n$ be independent random variables in a finite space $\Omega$, with distribution $P_{Q_i}$ respectively. Denote random vector $Q^n:=(Q_1, \cdots, Q_n)$ and the joint distribution $P_{Q^n}:=\Pi_{i=1}^n P_{Q_i}$. Suppose that event ${\calA}^{(n)} \in \Omega^n$ is such that $Pr(Q^n \in {\calA}^{(n)}) \geq 2^{-n c_n}$ for $c_n \geq 0$. Then for any $\lambda>1$, $P(Q^n \in \Gamma_{n \lambda \sqrt{c_n}} ({\calA}^{(n)})) \geq 1-1/\lambda$.
\end{lemma}
\vspace{2mm}

{\bf Proof:} The proof follows Marton's proof \cite{Marton1986} and the summary in El Gamal's slides \cite{ElGamal2010}. Please see the details in Appendix \ref{AppendixA}. \qed

As can be seen, the above two lemmas consider how large (in Hamming distance) one should blow-up an event set so that the larger set has a non-trivial probability.

Similar result is needed for the relay channel we consider.
Recall that node $X$ sends a code word $X^n$ uniformly picked from its code book ${\calC}^{(n)}$. This generates an observation $Z^n$ at node $Z$, which has a `color' $\hat{Z^n}$. As will be shown later, the conditional entropy $H(\hat{Z^n}|X^n)$ is a key parameter in  bounding the feasible rate away from the cut-set bound. Given $H(\hat{Z^n}|X^n) = n a_n$, we show that there exist a Hamming distance determined by $a_n$, a non-trivial set of codewords ${\calC}^{(n)}_1 \subseteq {\calC}^{(n)}$ , and a set of special colors associated with each such codeword satisfying the following. If a codeword $x^n$ from ${\calC}^{(n)}_1$ is sent, then for each special color $j$ of $x^n$, blow up the set of $z^n$'s of color $j$ by the distance specified. Then this new set has a non-trivial probability.
Specifically we have the following.
\vspace{2mm}
\begin{theorem}
\label{theorem_blowingupwithConditionalEntropy}
Assume that $H(\hat{Z^n}|X^n) = n a_n$. Then for any given $\lambda>1$, there exists a set of codewords ${\calC}^{(n)}_1$ satisfying the following:
\begin{itemize}
\item $Pr(X^n\in {\calC}^{(n)}_1) \geq 1-1/ \lambda$;
\item For each code word $x^n$ in $\calC^{(n)}_1$, there is a set of colors $S(x^n) \subseteq \{1, \cdots, 2^{n R_0}\}$ such that $Pr(\hat{Z^n} \in S(x^n) |X^n=x^n) \geq 1-1/\lambda$. Furthermore,
    for each $j$ of $S(x^n)$, we have
$$Pr \left(Z^n \in \Gamma_{n \lambda^{3/2} \sqrt{a_n}} (\calA_j^{(n)}) |x^n \right) \geq 1-1/\lambda>0,$$
where $\calA_j^{(n)} :=\{z^n \in \Omega_Z^n: \, \hat{z^n}=j\}$.
\end{itemize}
\end{theorem}

{\bf Proof:} Please see in Appendix \ref{AppendixA}. \qed

\section{Upperbound when $Y$ and $Z$ are conditionally I.I.D. given $X$}
In this section, we consider the case when $Y$ and $Z$ are conditionally i.i.d. given $X$. That is, $\Omega_Y = \Omega_Z :=\Omega$, and for all $\omega \in \Omega$ and $x\in \Omega_X$, $Pr(Y=\omega | X=x)$ equals $Pr(Z=\omega | X=x)$.
Two inequalities on any feasible rate are introduced, both taking $H(\hat{Z^n} |X^n)$ as parameter.

The first one is Fano's inequality as follows.
\begin{lemma}
\label{lemma6} {\em [Fano's Inequality]}
Denote $H(\hat{Z^n}|X^n)=n a_n$. For any feasible rate $R$, we have $R \leq C_{XY}+R_0-a_n+o(1),$ as $n \to \infty$.
\end{lemma}
Proof:
Since the code book is feasible, we have $H(X^n)=nR$ and, by Fano's lemma \cite{CoverThomas1991}, \\
$H(X^n|Y^n, \hat{Z^n})= n \cdot o(1)$. So
\begin{eqnarray}\label{Derivation1}
\nonumber
&& n(R+o(1))=I(X^n; Y^n, \hat{Z^n})  = I(X^n; Y^n) +H(\hat{Z^n}|Y^n)-H(\hat{Z^n}|X^n)\\
&& \leq n C_{XY}+nR_0 -H(\hat{Z^n}|X^n).
\end{eqnarray}
\qed

Now we introduce the following definition.
\begin{definition}
A {\em ball of radius $r$ centered at a point} $x_0^n$ in a  space $\Omega^n$ is denoted as $Ball_{x_0^n}(r)$, and is defined as the set of points in $\Omega^n$ that is within Hamming distance $r$ of $x_0^n$. When $r$ is not an integer, the minimum integer no less than $r$ is used instead. In the paper, we often use $Ball(r)$ when there is no confusion.
\end{definition}

The following is true on the volume of a ball -- the number of points enclosed.
\begin{remark} \label{remark_onballsize}
For fixed constant $\rho \in [0, 1]$, we have
$ |Ball(n \rho)| = {n \choose \rho n} |\Omega|^{\rho n}$.
By Lemma 17.5.1 in the 2006 edition of \cite{CoverThomas1991}, we have $\frac{1}{n} \log |Ball(n \rho)| =\rho \log |\Omega| + H_2(\rho) +o(1),$
where the $o(1)$ is only a function of $n$, and $H_2(\rho)$ is the binary entropy function $-\rho \log \rho - (1-\rho) \log (1-\rho)$.
\end{remark}

The second inequality is the following. It hinges on the fact that any decoding strategy {\em solely} based on $Y^n$ is subject to the inequality in Theorem \ref{Arimoto1973}. While, given a feasible strategy, upon which a procedure for node $Y$ to guess $X^n$ can be derived.
\vspace{2mm}
\begin{theorem}\label{theorem_main0}
Assume that $Y$ and $Z$ are i.i.d. given $X$, and $H(\hat{Z^n} |X^n) = n a_n$. Also assume that rate $R>C_{XY}$ is achievable. Then for all $\lambda>1$, there exist $\delta_n$ going to zero, determined by $n$ and $\lambda$ only, and integer $N_1$, such that for $n \geq N_1$,
\begin{eqnarray*}
\frac{1}{n} \log \left| Ball(n \lambda^{3/2} \sqrt{a_n}) \right| +\delta_n \geq \calE(R),
\end{eqnarray*}
where $\calE(R)$ is defined in (\ref{def_decoErrorProb}) for the $XY$ channel.
\end{theorem}

{\bf Proof:}
We present the main ideas here. The detailed proof is in Appendix \ref{AppendixB}.

By definition, for a feasible coding strategy, there associates a decoding function, $f_n(\hat{Z^n}, Y^n)$, at node $Y$ which correctly maps $(\hat{Z^n}, Y^n)$ to the codeword almost surely. So to construct a decoding strategy for node $Y$ to be depending on $Y^n$ only, one natural way is to let $Y$ guess the color $\hat{Z^n}$ and then apply $f_n(\cdot, \cdot)$.

The following strategy is proposed. Node $Y$ paints every point $\omega^n$ in $\Omega^n$ the same color node $Z$  would paint, namely $\hat{z}^n$. Once receiving $Y^n$, node $Y$ draws a Hamming ball of radius $n \lambda^{3/2} \sqrt{a_n}$ around $Y^n$ in $\Omega^n$. Then it randomly and uniformly picks a point in the ball and finds its color as a guess on $\hat{Z^n}$.

We now show that the probability of guessing $\hat{Z^n}$ correctly this way is about order $c_1 \frac{1}{|Ball(n \lambda^{3/2} \sqrt{a_n})|}$, with $c_1>0$ being constant. Note that if this is true, then the theorem is immediate by applying Theorem \ref{Arimoto1973}.

Actually, by Theorem \ref{theorem_blowingupwithConditionalEntropy}, for a probability $p_1>0$, $Z^n$'s color $\hat{Z^n}$ is from a special color set $S(X^n)$ of the transmitted codeword $X^n$.
For each such special color, say $j$, blowing up all the points in $\Omega^n$ of color $j$ by Hamming distance $n \lambda^{3/2} \sqrt{a_n}$ results in a set with the following property. If one generates a random variable based on $Z^n$'s distribution given $X^n$, then this random variable will be in this set with probability no less than $p_1$. Since $Y^n$ is such a random variable, $Y^n$ is within distance $n \lambda^{3/2} \sqrt{a_n}$ of a $\hat{Z^n}$-colored point with probability no less than $p_1^2$. Thus overall, the probability that our strategy guesses $\hat{Z^n}$ correctly -- equivalent to guessing $X^n$ -- is no less than $c_1 \frac{1}{|Ball(n \lambda^{3/2} \sqrt{a_n})|}$, with $c_1>0$ being a function of $p_1$. \qed

Combining Lemma \ref{lemma6} and Theorem \ref{theorem_main0} gives the following main theorem.
\begin{theorem}
\label{theorem_main1}
Assume that $Y$ and $Z$ are i.i.d. given $X$. Then there exists $a \in [0, R_0]$ such that any feasible rate $R$ larger than $C_{XY}$ satisfies:
$R - C_{XY} \leq R_0 - a$ and $\calE(R) \leq H_2(\sqrt{a}) +\sqrt{a} \log |\Omega|$.
\end{theorem}

Proof: Assume that $H(\hat{Z^n} | X^n)/n = a_n$. From Lemma \ref{lemma6}, Theorem \ref{theorem_main0}  and Remark \ref{remark_onballsize}, we know
$R-C_{XY} \leq R_0-a_n+o(1)$ and $\calE(R) \leq H_2(\lambda^{3/2} \sqrt{a_n})+ \lambda^{3/2} \sqrt{a_n} \log|\Omega| +o(1)$.
Suppose $\limsup a_n = a$, which exists because $a_n$ is finite in $[0, R_0]$. Then
$R-C_{XY} \leq R_0-a$ and $\calE(R) \leq H_2(\lambda^{3/2} \sqrt{a})+ \lambda^{3/2} \sqrt{a} \log|\Omega| $. Because this is valid for any $\lambda>1$, we know the theorem is true.\qed

The following is immediate by Remark \ref{remarkOnerrexponent} and that $H_2( \sqrt{a})+  \sqrt{a} \log|\Omega|$ is continuous in $a$ and is zero at $a=0$.
\begin{corollary}
\label{corollary_main1}
When $Y$ and $Z$ are i.i.d. given $X$, and $R>C_{XY}$ is feasible, then $R$ is strictly less than $C_{XY}+R_0$.
\end{corollary}

Now we take the binary erasure channel (BEC) as an example for detailed analysis.

{\bf Example: Detailed Analysis on the BEC.}
Suppose both $XY$ and $XZ$ are conditionally i.i.d. binary erasure channels with erasure probability $\epsilon$, as defined by
$Pr(y=x|x)=1-\epsilon, \,\, Pr(y=erasure|x)=\epsilon, \,\, \forall x\in \{0, 1\}$.

The corresponding $\calE(R)$ can be determined as follows. The detailed derivation is in Appendix \ref{AppendixB}.
\begin{eqnarray}
\label{ERforBEC}
\calE(R) = \left\{
\begin{array}{l}
R \log \frac{R \epsilon}{(1-\epsilon) (1-R)} -\log \left(\frac{R \epsilon}{1-R} +\epsilon \right),
 \,\, R \in  (1-\epsilon, 1-\frac{\epsilon}{2-\epsilon});\\
R- \log (2-\epsilon), \,\, R \geq 1-\frac{\epsilon}{2-\epsilon}.
\end{array}
\right.
\end{eqnarray}

With this, Theorem \ref{theorem_main1} can be applied to find the bound numerically on the achievable rate for any given $R_0$.
The following is a plot for the case when $\epsilon=0.5$. The bound is nevertheless very close to the cut-set bound.


\begin{figure}[htbp]
  \centering
    \includegraphics[height=12cm, width=16cm]{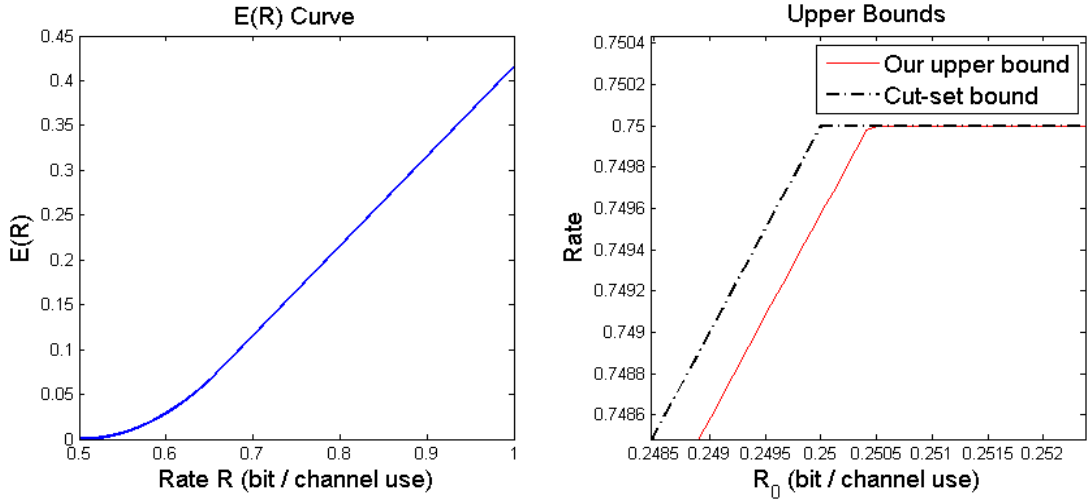}
    \caption{Numerical result on the BEC channel when $Pr(erasure)=0.5$. Note that $C_{XY}=C_{XZ}=0.5$, while the capacity between $X$ and $(Y, Z)$ is $1-0.5^2=0.75$.}
   \label{Fig:BECplots}
\end{figure}

\section{When the Channel is Statistically Degraded}

In this section, we extend the result in the previous section to the case when the channel is statistically degraded. We say $Z$ is a statistically degraded version of $Y$ with respect to $X$ if there exists a transition probability distribution $q_1(z|y)$ such that $p(z|x)=\sum_y q_1(z|y) p(y|x)$. Accordingly we say that channel $XYZ$ is degraded. Similarly, $Y$ is a statistically degraded version of $Z$ with respect to $X$ if there exists a probability distribution $q_2(y|z)$ such that $p(y|x)=\sum_z q_2(y|z) p(z|x)$. In this case, channel $XZY$ is degraded.
Note that \cite{Zhang1988} considers the case when $XZY$ is statistically degraded.

\subsection{When $XYZ$ is Statistically Degraded}
\begin{figure}[htbp]
  \centering
    \includegraphics[height=4.5cm]{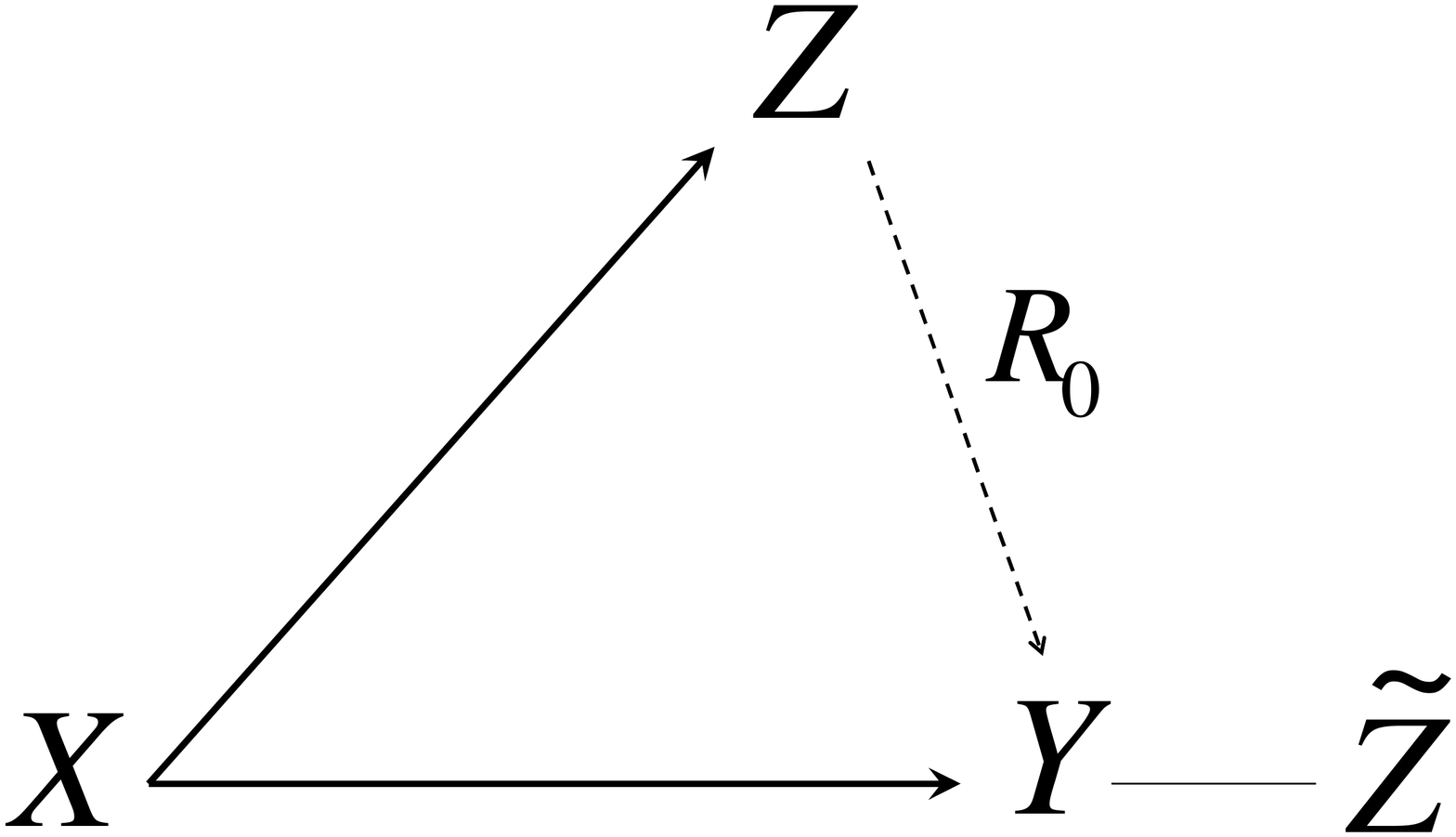}
    \caption{Augmented Network when $XYZ$ is degraded. $\tilde{Z}$ is generated based on $q_1(z|y)$.}
    \label{Fig:augmentedFigXYZ}
\end{figure}

The following procedure can be employed by $Y$ to decode $X^n$ solely based on observation $Y^n$. At $i$-th transmission, upon receiving an observation $Y_i$, it generates a random variable $\tilde{Z_i}$ based on the transition probability $q_1(z|y)$; thus
for the observed $Y^n$, a $\tilde{Z}^n$ is generated. Now consider the relay channel formed by $X, \tilde{Z},$ and $Z$; see Figure \ref{Fig:augmentedFigXYZ}. It is obvious that $Z$ and $\tilde{Z}$ are i.i.d. given $X$.
The same procedure in Section IV, namely the method for $\tilde{Z}$ (it is actually node $Y$) to guess $\hat{Z}^n$ and the derivation on the decoding probability, can be applied.
This leads to the following, similar to Theorems \ref{theorem_main0} and \ref{theorem_main1}.

\begin{theorem}\label{theorem_markov11}
Suppose that $XYZ$ is statistically degraded.
Denote $H(\hat{Z^n} |X^n) = n a_n$. Then for all $\lambda>1$, there exists $\delta_n \to 0$, determined by $n$ and $\lambda$ only, such that
 $\frac{1}{n} \log|Ball_{\Omega_Z}(n \lambda^{3/2} \sqrt{a_n})| +\delta_n \geq \calE_Y(R),$ for $R>C_{XY}$. Here $\calE_Y(R)$ is  as defined in (\ref{def_decoErrorProb}) for the $XY$ channel.
\end{theorem}

\begin{theorem}
\label{theorem_markov12}
Suppose that $XYZ$ is statistically degraded. Then there exists $a \in [0, R_0]$ such that any achievable rate $R$ larger than $C_{XY}$ satisfies:
$R - C_{XY} \leq R_0 - a$ and $\calE_Y(R) \leq H_2(\sqrt{a}) +\sqrt{a} \log |\Omega_Z|$.
\end{theorem}

\subsection{When $XZY$ is Statistically Degraded}
The upper bound for this case can be derived by considering the decoding probability when node $Z$ tries to decode $X^n$ solely based on $Z^n$ as follows.


\begin{figure}[htbp]
  \centering
    \includegraphics[height=4.5cm]{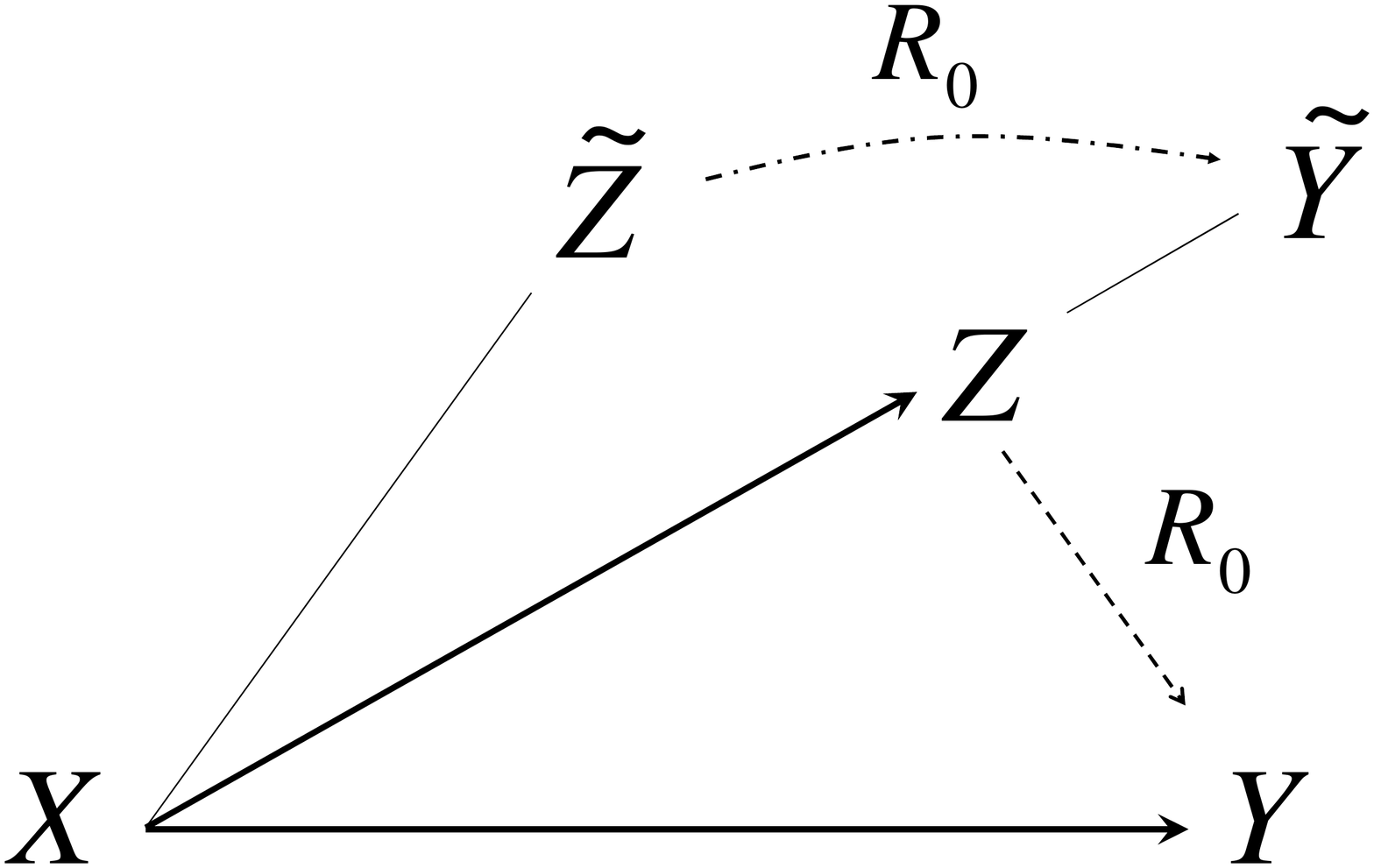}
    \caption{Augmented Network when $XZY$ is degraded. Node $Z$ now tries to decode $X^n$ solely based on $Z^n$. $\tilde{Y}$ is generated based on $q_2(y|z)$. $\tilde{Z}$ is a random variable with the same distribution of $Z$ given $X$.}
    \label{Fig:system2}
\end{figure}

Build a new channel based on the relay channel $XYZ$ as depicted in Figure \ref{Fig:system2}.
First, add a new random variable $\tilde{Z}$ which is independent of others given $X$ and has the same distribution as $Z$ given $X$. Then add another random variable $\tilde{Y}$ based on $Z$ as follows. Whenever $Z$ is received, node $Z$ generates $\tilde{Y}$ according to $q_2(y|z)$. Thus we have a new channel $X Z \tilde{Y} \tilde{Z}$. Finally add a lossless link of rate $R_0$ from $\tilde{Z}$ to $\tilde{Y}$.
Since the channels $X \tilde{Y} \tilde{Z}$ and $XYZ$ are equivalent statistically, any rate achievable by the $XYZ$ channel must be achievable by the channel $X Z \tilde{Y} \tilde{Z}$. Here $(Z, \tilde{Y})$ is considered as one single node. To see this, given observation $\tilde{Z}^n$, node $\tilde{Z}$ maps it to a color $\hat{ \tilde{Z}}^n$ based on the same mapping from $Z^n$ to $\hat{Z^n}$.
For any feasible coding strategy, node $Z$ invokes the associated decoding function $f_n(\hat{ \tilde{Z}}^n, \tilde{Y}^n)$ to decode $X^n$.

Now consider the channel $X Z \tilde{Y} \tilde{Z}$. Node $Z$ can guess $X^n$ based solely on $Z^n$ by the following procedure. Assume $H(\hat{ \tilde{Z}}^n |X^n) = n a_n$, and fix a constant $\lambda>1$. Node $Z$ draws a ball of radius $n \lambda^{3/2} \sqrt{a_n}$ around $Z^n$. Because $Z$ and $\tilde{Z}$ are i.i.d. given $X$,  as shown in the proof for Theorem \ref{theorem_main1}, the color $\hat{ \tilde{Z}}^n$ is contained in the ball with non-diminishing probability.
Randomly pick a point $\omega^n$ in the ball, node $Z$ announces $f_n(\hat{ \omega^n}, \tilde{Y}^n)$ as the code word.
By similar argument in the previous section, the following is true.
\begin{theorem}\label{theorem_markov21}
Assume $XZY$ is statistically degraded. Denote $H(\hat{Z^n} |X^n) = n a_n$. Then for all $\lambda>1$, there exists $\delta_n$ going to zero, determined by $n$ and $\lambda$ only, such that
$$\frac{1}{n} \log|Ball_{\Omega_Z}(n \lambda^{3/2} \sqrt{a_n})| +\delta_n \geq \calE_Z(R).$$
\end{theorem}

\begin{theorem}
\label{theorem_markov22}
Assume $XZY$ is statistically degraded. Then there exists $a \in [0, R_0]$ such that any achievable rate $R$ larger than $C_{XZ}$ satisfies:
$R - C_{XY} \leq R_0 - a$ and $\calE_Z(R) \leq H_2(\sqrt{a}) +\sqrt{a} \log |\Omega_Z|$.
\end{theorem}

\section{Channel simulation with side information}

In the previous two sections, the new inequality is based on decoding error probability. The key step is for a node (e.g. $Y$) to guess the color of another node's observation (e.g. $Z^n$) by generating a random variable with the same distribution given $X^n$.
This is readily doable when the channel is statistically degraded. For general cases, one needs new method. In this and the next sections, we show that this can be done by generalized results from channel simulation. To the best knowledge of the author, this is the first time channel simulation is applied in analyzing the relay channel capacity. For a clear presentation, we first introduce channel simulation and generalize a basic result in this section suitable for our purpose. In the next section, the result will be applied to bound the relay channel capacity.

\subsection{Channel Simulation and its Adaptation for the Relay Channel Considered}

\begin{figure}[htbp]
  \centering
    \includegraphics[height=7cm]{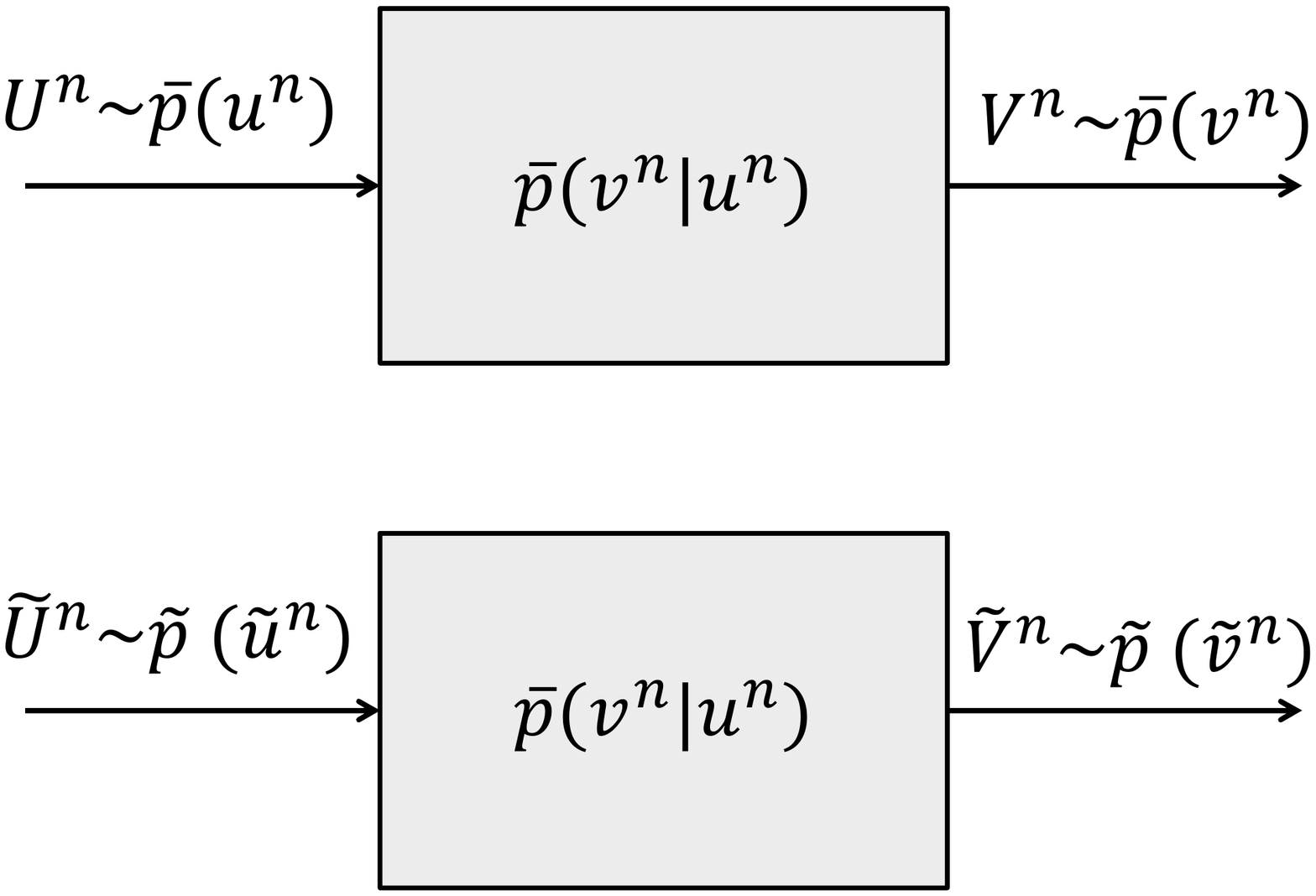}
    \caption{Channel simulation in its original formulation (CS-Basic). Top: The original channel to be simulated; Bottom: Simulated channel.}
    \label{Fig:channelSimu1}
\end{figure}

{\bf Channel simulation (CS-Basic).} In its original formulation, {\em channel simulation} (e.g. \cite{HanVerdu1993} \cite{Wyner1975}) concerns the following problem in general. Suppose there is a source $U^n$, randomly generated according to distribution $\bar{p}(u^n)$, and a channel defined by $\bar{p}(v^n|u^n)$; see Figure \ref{Fig:channelSimu1}. The channel output is denoted as $V^n$ with distribution $\bar{p}(v^n)$. Then the task of channel simulation is to efficiently design a $\tilde{U}^n$ with certain cardinality and an associated distribution $\tilde{p}(\tilde{u}^n)$ such that, when one inputs the channel based on $\tilde{p}(\tilde{u}^n)$, the induced output distribution $\tilde{p}(\tilde{v}^n)$ is close to $\bar{p}(v^n)$ in the sense that
$$d(V^n, \tilde{V}^n):=\sum_{v^n} | \bar{p}(v^n)- \tilde{p}(v^n)| \to 0.$$
The optimization focuses on minimizing
the cardinality of the support of $\tilde{p}(\tilde{u}^n)$. Note that $d(V^n, \tilde{V}^n)$ also equals $\max_A \frac{1}{2} |Pr(V^n \in A)-Pr(\tilde{V}^n \in A)|$.


\par

For bounding the capacity of the relay channel under consideration, we adapt the channel simulation formulation to the following.

\begin{figure}[htbp]
  \centering
    \includegraphics[height=7cm]{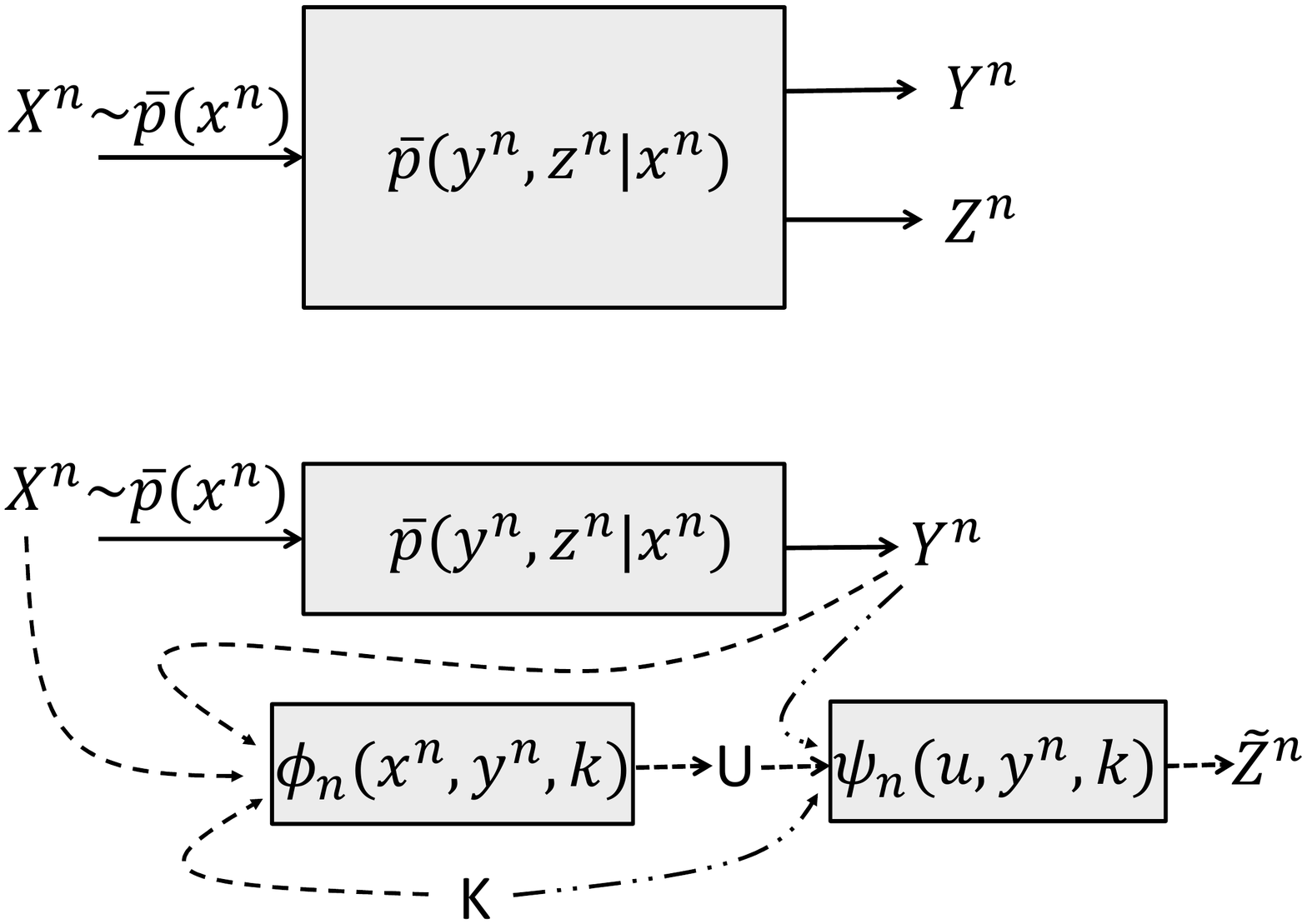}
    \caption{Channel simulation with side information and common randomness (CS-SICR). Top: The channel to be simulated. Bottom: Given $X^n$, $Y^n$ and $K$, channel encoder applies $\phi_n$ to generate code $U$; channel decoder applies $\psi_n$ to generate $\tilde{Z}^n$, which simulates $Z^n$.}
    \label{Fig:channelSimu2}
\end{figure}

{\bf Channel Simulation with Side Information and Common Randomness (CS-SICR).}
The following channel is considered, as shown in Figure \ref{Fig:channelSimu2}.
The source node $X$ produces $X^n$,  which is generated from a code book $\calC^{(n)}=\{c_1, \cdots, c_M\}$ with probability distribution $\bar{p}(X^n=c_j)=\frac{1}{M}$ for all $j$. The channel output is $Z^n$. Moreover, there is also a random variable $Y^n$ as side information. The channel is defined by $\bar{p}(y^n, z^n | x^n)$, and the random variables have a joint distribution $\bar{p}(x^n, y^n, z^n)$.

The channel simulation procedure is as follows.
A ``channel encoder" sees the source $X^n$, side information $Y^n$, as well as a ``common" random variable $K$ which is uniformly distributed on $\{1, 2, \cdots, 2^{nR_2} \}$, where $R_2$ is a constant. It determines a (simulation) code word $U \in \{1, 2, \cdots, 2^{nR_1} \}$ based on an encoding function $\phi_n(x^n, y^n, k)$, which is a probability distribution on $\{1, 2, \cdots, 2^{nR_1} \}$.
There is a ``channel decoder" which also observes $Y^n$ and $K$. Upon receiving $U$, it will generate an output random variable
$\tilde{Z}^n$ based on a function $\psi_n(u, y^n, k)$. Suppose the joint distribution among the random variables is $q(x^n, y^n, z^n, u, k)$. The objective of the channel simulation is to design $\phi_n(\cdot, \cdot, \cdot)$ and $\psi_n(\cdot, \cdot, \cdot)$ such that
$$\sum_{x^n, y^n, z^n} \left| \bar{p}(x^n, y^n, z^n) - \bar{p}(x^n, y^n) Q(z^n|x^n, y^n) \right| \to 0,$$
 where $Q(z^n | x^n, y^n)$ is the conditional distribution induced from the joint distribution $q(x^n, y^n, z^n, u, k)$.


\par

\begin{remark}
\label{remark_diff2CuffFormulation}
Note that the above formulation is based on \cite{Cuff2008}. Compared to \cite{Cuff2008}, there are two differences. First, the source $X^n$ is not generated from i.i.d. random variables $X_1, X_2, \cdots, X_n$ based on a distribution $p(x)$. Instead, here the source is uniformly picked from a code book. Secondly, there is a side information $Y^n$ in our formulation.
\end{remark}

\subsection{Why CS-SICR Can Be Used for Bounding the Capacity of the Relay Channel}
Before going to deriving results for the special channel simulation, we first briefly explain why the seemingly irrelevant channel simulation can be applied towards bounding the capacity of the relay channel. Suppose such simulation procedure has been established by designing $\phi_n(\cdot, \cdot, \cdot)$ and $\psi_n(\cdot, \cdot, \cdot)$. Then if $U$ were given, node $Y$ in our relay channel would be able to use $\psi_n(u, y^n, k)$ to generate a random variable of the same distribution\footnote{The exact meaning will be made clear later.} of $Z^n$ given $X^n$. This is because $Y$ knows $Y^n$ and the common randomness $K$. Thereafter, one follows the procedures in the previous sections for node $Y$ to guess the code word $X^n$ and thus leads to the new inequality. However, here $U$ is an unknown element in $\{1, 2, \cdots, 2^{nR_1} \}$. Thus our new guessing strategy starts by first picking a random element in $\{1, 2, \cdots, 2^{nR_1} \}$ as a guess on $U$.

Based on this thinking, the optimization on the channel simulation is to minimize $R_1$.

\subsection{Results on Channel Simulation}

A few definitions need to be introduced.
\begin{definition} For a pair of random variables $U^n$ and $V^n$ with joint distribution $\bar{p}(u^n, v^n)$, the {\em point mutual information} $i(U^n; V^n)$ is defined as the random variable $\log \frac{\bar{p}(V^n|U^n)}{\bar{p}(V^n)}$. Note that $I(U^n; V^n) = E i(U^n; V^n)$. Similarly, when there exists another random variable $Y^n$, define {\em conditional point mutual information } $i(U^n; V^n |Y^n)$ as $\log \frac{\bar{p}(V^n|U^n, Y^n)}{\bar{p}(V^n |Y^n)}$.
\end{definition}

\begin{definition}
The {\em limsup in probability} of a sequence of random variables $\{T_n\}$ is the smallest $\beta$ such that for all $\epsilon>0$, $\lim_n Pr(T_n \geq \beta+\epsilon)=0$.
 The {\em liminf in probability} of $\{T_n\}$ is the largest $\alpha$ such that for all $\epsilon>0$, $\lim_n Pr(T_n \leq \alpha-\epsilon)=0$.
\end{definition}

A basic result of channel simulation is the following lemma. It shows that for the basic channel simulation (CS-Basic), the limsup of the average point mutual information is the rate required.

\begin{lemma}
 \label{Lemma_HanVerdu} {\em [Section IV around Equation (4.1) in \cite{HanVerdu1993}]}\footnote{The concept was first introduced in \cite{Wyner1975}.} In CS-Basic, assume that random variables $U^n$ and $V^n$ have joint distribution and marginal distributions $\bar{p}(u^n, v^n)$, $\bar{p}(u^n)$ and $\bar{p}(v^n)$, respectively.  Define $\bar{I}(U; V):=\limsup \frac{1}{n} \log \frac{\bar{p}(V^n|U^n)}{\bar{p}(V^n)}$ (in probability). For a given $\gamma>0$, generate $M=2^{n \bar{I}(U; V)+n \gamma}$ i.i.d. random variables  $\tilde{U}^n_j$, $j=1, \cdots, M$, according to $\bar{p}(u^n)$. Assume $\tilde{U}^n_j=c_j$, $j=1, \cdots, M.$ Define an associated distribution
$P_{\tilde{V}^n[c_1, \cdots, c_M]}(v^n) := \frac{1}{M} \sum_{j=1}^M \bar{p}(v^n|c_j).$
Then $\lim_n E d(V^n, \tilde{V}^n[c_1, \cdots, c_M]) =0.$
\end{lemma}

For channel simulation CS-SICR, not surprisingly, the limsup of the average conditional point mutual information is the rate needed.
\begin{theorem} \label{theorem_simuSideInfo}
Consider the channel simulation problem CS-SICR with side information $Y^n$ and common randomness $K$. For any $\delta>0$, there exist channel simulation encoding $\phi_n(x^n, y^n, k)$ and decoding $\psi_n(u, y^n, k)$ with rate $R_1=\limsup i(X^n; Z^n | Y^n) / n + \delta$  and $R_2$ sufficiently large such that
\begin{eqnarray}
\label{formula_theorem10}
\sum_{x^n, y^n, z^n} \left| \bar{p}(x^n, y^n, z^n) - \bar{p}(x^n, y^n) Q(z^n|x^n, y^n) \right| \to 0, \, n \to \infty,
\end{eqnarray}
where $Q(z^n|x^n, y^n)$ is the conditional distribution induced from the joint distribution $q(x^n, y^n, z^n, u, k)$.
\end{theorem}

Proof:  The theorem is a generalization to result in \cite{Cuff2008}. The details can be found in Appendix \ref{AppendixC}. \qed

\section{A general upper bound based on channel simulation with side information}
Given the preparation in the previous section, we are now ready to present a general upper bound for the relay channel under consideration.
We use ``the relay channel" to refer to the channel we are considering in Figure \ref{Fig:system0} and defined in Section II.

First we introduce a {\em companion channel} to the relay channel.
\begin{definition}
Suppose the relay channel in Figure 1 is defined by a conditional distribution $p(y, z|x)=p(y|x) p(z|x)$. And $\calC^{(n)} :=\{c_1, \cdots, c_M\}$ is a code book for a feasible coding strategy. A {\em companion (simulated) channel} is a memoryless channel defined by a conditional distribution $\bar{p}(y, z|x)$ which satisfies the following:
\begin{enumerate}
\item[i)] For all $x$, $y$ and $z$, $\bar{p}(y|x)=p(y|x)$, and $\bar{p}(z|x)=p(z|x)$. That is, the marginals are the same.
\item[ii)] The input distribution is $\bar{p}(X^n=c)=1/M$, for all $c \in \calC^{(n)}$.
\end{enumerate}
Furthermore, we use $\bar{C}_{XY}$ and $\bar{C}_{XYZ}$ to denote the capacities between $X$-$Y$ and $X$-$(Y, Z)$, respectively, in this {\em companion channel}.
\end{definition}

One notices immediately that $C_{XY}=\bar{C}_{XY}$.
\begin{remark}
For the relay channel, if there exists $q_1(z|y)$ such that $p(z|x)=\sum_y q_1(z|y) p(y|x)$, one can choose $\bar{p}(y, z|x)=p(y|x) q_1(z|y)$. In the case when $Y$ and $Z$ are i.i.d. given $X$, it leads to   $\bar{p}(y, z|x) = p(y|x) \cdot 1_{[z=y]}$.
\end{remark}

For the companion channel, we have the following two lemmas relating point mutual information to channel capacity.

\begin{lemma} \label{lemma_fromixy2ixzgivenY}
If $\liminf \frac{1}{n} i(X^n; Y^n)= \bar{C}_{XY}-c_1$, with $c_1 \geq 0$, then
$$\limsup \frac{1}{n} i(X^n; Z^n |Y^n) \leq \bar{C}_{XYZ}-\bar{C}_{XY}+c_1.$$
\end{lemma}

Proof: Since the channel is memoryless, finite and discrete, the communication channel between $X$ and $(Y, Z)$ satisfies the strong converse property. By Lemma 10 in \cite{HanVerdu1993}, we know that $\limsup \frac{1}{n} i(X^n; Y^n, Z^n) \leq \bar{C}_{XYZ}$ . Since
$i(X^n; Y^n, Z^n)=i(X^n; Y^n)+i(X^n; Z^n | Y^n),$
the conclusion is obvious. \qed

\begin{lemma} \label{lemma_i2Irelation}
Suppose the code book size is $2^{nR}$ in the relay channel.
If $\liminf \frac{1}{n} i(X^n; Y^n) \geq \bar{C}_{XY}-c_1$ for $c_1\geq 0$, then there exists $c_2\geq 0$ and $n_k$ going to infinity such that:
\begin{enumerate}
\item[i)] $\lim_k \frac{1}{n_k} I(X^{n_k}; Y^{n_k}) \leq \bar{C}_{XY}-c_2$; and
\item[ii)] $c_2$ is positive if $c_1$ is positive.
\end{enumerate}
\end{lemma}

Proof: Please see in Appendix \ref{AppendixD}. \qed

Now we present the following main result which is a generalization to Theorem \ref{theorem_markov12} in Section V.
\begin{theorem}
\label{theorem_mainGeneral}
Suppose code book $\calC^{(n)}$ of rate $R$ is feasible for the relay channel, and $\liminf i(X^n; Y^n)/n = C_{XY}-c_1$. Then for any companion channel $\bar{p}(y, z|x)$, there exist constants $c_2 \geq 0$ and $a \geq 0$ such that:
\begin{enumerate}
\item [i)] $R \leq C_{XY}-c_2 +R_0-a$;
\item [ii)] $\calE_Y(R) \leq \bar{C}_{XYZ}-C_{XY}+c_1 +H_2(\sqrt{a}) + \sqrt{a} \log |\Omega_Z|$;
\item [iii)] $c_2$ is positive when $c_1>0$, as identified in Lemma \ref{lemma_i2Irelation}.
\end{enumerate}
\end{theorem}

Proof:
We scatch the main ideas here. The detailed proof is in Appendix \ref{AppendixD}.

i) and iii) are due to Fano's lemma and Lemma \ref{lemma_i2Irelation} as follows.
Denote $a_n:=H(\hat{Z^n}|X^n) / n$ for the relay channel.
For any $n$, we know $H(X^n)=nR$ and, by Fano's lemma, $H(X^n|Y^n, \hat{Z^n})= n \cdot o(1)$. Similarly as before, this leads to
$$ R\leq I(X^n; Y^n)/n +R_0 - a_n +o(1).$$
Thus, since $\liminf i(X^n; Y^n)/n = C_{XY}-c_1$, by Lemma \ref{lemma_i2Irelation}, we know there exits $n_k \to \infty$ such that
$
R\leq C_{XY}-c_2 +R_0 - a_{n_k} +o(1).
$
Denoting $a:=\limsup a_{n_k}$, we have
$
R\leq C_{XY}-c_2 +R_0 - a.
$
This gives i) and iii).

ii) can be shown by applying channel simulation results for the companion channel $\bar{p}(y, z|x)$.

By Lemma \ref{lemma_fromixy2ixzgivenY} and Theorem \ref{theorem_simuSideInfo}, for any $\delta>0$, with rate
$
R_1:= \bar{C}_{XYZ}-\bar{C}_{XY}+c_1+\delta,
$
 one can simulate the channel $\bar{p}(x^n, y^n, z^n)$ based on side information $Y^n$ and a common randomness $K$.
This involves constructing $\phi_n(x^n, y^n, k)$ and $\psi_n(u, y^n, k)$.
In the relay channel, node $Y$ can utilize this to produce a $\tilde{Z^n}$ with distribution close to that of the relay's observation $Z^n$ as follows. To generate a channel simulation output based on $\psi_n(\cdot, \cdot, \cdot)$, it needs $U$, $K$, and $Y^n$.
It has  $Y^n$ and $K$. For $U$, there are total $2^{n R_1}$ possibilities. Node $Y$ picks an element  $\tilde{U}$ uniformly in $\{1, 2, \cdots, 2^{nR_1} \}$ as $U$, and generates a $\tilde{Z}^n$ based on $\psi_n(\tilde{U}, Y^n, K)$. Note that the probability to hit the correct one, i.e. $\tilde{U}=U$, is at least $2^{-n R_1}$.

Given $\tilde{Z}^n$, node $Y$ can apply the same procedure and argument as in Section IV to guess $X^n$. Specifically, it draws a ball of radius $n \lambda^{3/2} \sqrt{a_n}$ around $\tilde{Z^n}$ in the space $\Omega_Z^n$, for a constant
$\lambda>1$. Then it picks a point $\omega^n$ uniformly in the ball and applies the known decoding function $f_n(\hat{\omega}^n, Y^n)$ to guess $X^n$, where $\hat{\omega}^n$ is the `color' of $\omega^n$.

Now we analyze the decoding probability of the above procedure.
The decoding would be successful if both the following conditions are true. First, node $Y$ chooses the correct $\tilde{U}$ to simulate the correct $\tilde{Z}^n$, i.e. $\tilde{U}=U$. Second, given a correct $\tilde{Z}^n$, node $Y$ hits the correct color in the ball of radius $n \lambda^{3/2} \sqrt{a_n}$ around $\tilde{Z^n}$. We hence have
 $$Pr(\mbox{Node $Y$ can decode correctly}) \geq  2^{-n R_1} \cdot  \frac{\mu_1}{|Ball(n \lambda^{3/2} \sqrt{ a_n})|},$$
where $\mu_1>0$ is a constant.

Based on the result of Arimoto \cite{Arimoto1973}, one must have
$$\calE_Y(R) \leq R_1 +\limsup \frac{1}{n} \log \left| Ball(n \lambda^{3/2} \sqrt{ a_{n}}) \right|.$$
Plugging in the bound on the ball's volume as in Remark \ref{remark_onballsize}, the above inequality leads to the desired claim ii).
\qed

\subsection{Discussion on Theorem \ref{theorem_mainGeneral}}
When $XYZ$ is statistically degraded, one can choose the companion channel such that $\bar{C}_{XYZ}=C_{XY}$, and $c_1=c_2=0$. To be more specific, one can make $XYZ$ to be a Markov chain. This shows that the bound, when $XYZ$ is degraded\footnote{A similar result to Theorem \ref{theorem_mainGeneral} can be derived when node $Z$ simulates $Y^n$. This will include the case when $XZY$ is statistically degraded.}, is a special case of Theorem \ref{theorem_mainGeneral}.

There are certainly cases where $\bar{C}_{XYZ}>C_{XY}$ no matter how one chooses the companion channel. In these cases, by purely looking at Theorem \ref{theorem_mainGeneral}, one can choose $a=0$ and $c_1=c_2=0$ without violating either i) or ii) for $R$ slightly larger than $C_{XY}$ (i.e. $R_0$ is close to zero). The effective bound becomes i), which is the same as the cut-set bound in this regime. When $R_0$ gets larger, the inequality in ii) becomes the effective bound. At this moment, our new bound deviates from the cut-set bound, and is strictly better.


\section{Concluding Remarks}
The paper presents a new technique for upper-bounding the capacity of the relay channel. Bound strictly better than the cut-set bound is achieved. One of the essential ideas is to let one node simulate the other node's observation.

However, requiring a lossless link between the relay and the destination makes it quite different than the original relay channel in \cite{Meulen1971}. It remains unclear how fundamental this requirement is to the new bounding method.

Interestingly, it is in general possible that the cut-set bound is tight even when the encoding rate is larger than the capacities of both $XY$ and $XZ$ channels.
For example, consider the following deterministic relay channel\footnote{This can be considered as a special case of \cite{Kim2008} with specific code design.}:
\begin{itemize}
\item
$\Omega_X=\{1, 2, 3, 4\}$, $\Omega_Y = \{'A', 'B'\}$, and $\Omega_Z=\{'C', 'D'\}$;
\item
$Y='A'$, for all  $X \in \{1, 2\}$; $Y='B'$ for all $X \in \{3, 4\}$;
\item
$Z='C'$, for all  $X \in \{1, 3\}$; $Z='D'$ for all $X \in \{2, 4\}$;
\item
There is a lossless link of rate $R_0$ from $Z$ to $Y$.
\end{itemize}
Note that $C_{XY}=C_{XZ}=1$. For this channel, the following strategy can send $1+R_0$ bit per channel use from $X$ to $Y$ when $R_0<1$. First, construct a code book $\calC_1$ of rate 1 based on hypothetical symbols $\{a, b\}$. Denote it as $\calC_1:=\{ \alpha^n(w_1): w_1=1, \cdots, 2^n\}$. Then construct a code book $\calC_2$ of rate $R_0$ based on hypothetical symbols $\{c, d\}$. Denote it as $\calC_2:=\{ \beta^n(w_2): w_2=1, \cdots, 2^{n R_0}\}$. To send a message $(w_1, w_2)$, node $X$ compares $\alpha^n(w_1)$ and $\beta^n(w_2)$, and produces a codeword as follows.
For each position $k$,
$$
x_k=\left\{
\begin{array}{l}
1, \mbox{ if } \alpha_k^n(w_1)=a, \beta_k^n(w_2)=c; \\
2, \mbox{ if }  \alpha_k^n(w_1)=a, \beta_k^n(w_2)=d; \\
3, \mbox{ if }  \alpha_k^n(w_1)=b, \beta_k^n(w_2)=c; \\
4, \mbox{ if }  \alpha_k^n(w_1)=b, \beta_k^n(w_2)=d; \\
\end{array}
\right.
$$
It is easy to check that $w_1$ can be decoded by node $Y$ and $w_2$ by node $Z$. Then node $Z$ can forward this message to $Y$.


\appendices

\section{} \label{AppendixA}

\subsection{Proof for Lemma \ref{BlowupLemma}}
The proof follows Marton's proof \cite{Marton1986} and the summary in El Gamal's slides \cite{ElGamal2010}.

The following lemma is from \cite{Marton1986}. Recall that the KL-divergence between two distributions $P_1$ and $P_2$ is defined as $D(P_1||P_2):=\sum_i P_1(i) \log(P_1(i)/P_2(i))$.

{\em Lemma 1 of \cite{Marton1986}:}
Let $Q^n := (Q_1, Q_2, \cdots, Q_n)$ and $\bar{Q}^n := (\bar{Q}_1, \bar{Q}_2, \cdots, \bar{Q}_n)$ be two series of random variables defined in $(\Omega, \cal{F})$. Let $Q_1, \cdots, Q_n$ be independent, each with  distribution $P_{Q_i}$ respectively. Denote $Q^n$'s joint distribution as $P_{Q^n}=\prod_{i=1}^n P_{Q_i}$ and  $\bar{Q}^n$'s distribution as $P_{\bar{Q}^n}$. Then, there exists a joint probability distribution $P_{Q^n, \bar{Q}^n}$ with these given marginals such that
$$\frac{1}{n} E(d_H (Q^n, \bar{Q}^n))
\leq \left( \frac{1}{n} D \left( P_{\bar{Q}^n} || \prod_{i=1}^n P_{Q_i} \right) \right)^{1/2}.$$

Now define
\begin{eqnarray*}
P_{\bar{Q}^n}(x^n) :=
\left\{
\begin{array}{l}
P_{Q^n |{\calA}^{(n)}} (x^n) = P_{Q^n}(x^n) / P_{Q^n} ({\calA}^{(n)}), \,\, \forall x^n \in {\calA}^{(n)}; \\
0, \quad \mbox{otherwise.}
\end{array}
\right.
\end{eqnarray*}
Then,
$D(P_{\bar{Q}^n} || \prod_{i=1}^n P_{Q_i}) = - \log P_{Q^n} ({\calA}^{(n)}) \leq n c_n.$

By the lemma, we know there exists a joint distribution such that $Ed_H(Q^n, \bar{Q}^n) \leq n \sqrt{c_n}$.

By the Markov inequality, for any $\delta>0$,
$$P_{Q^n, \bar{Q}^n} \left\{ d_H(Q^n, \bar{Q}^n) \leq n \delta_n \right\} \geq  1- \frac{\sqrt{c_n}}{\delta_n}.$$
If we choose $\delta_n=\lambda \sqrt{c_n} $ with $\lambda>1$, we therefore have
\begin{eqnarray}
\nonumber
&& P_{Q^n} (\Gamma_{n \delta_n} (\calA^{(n)})) =
P_{Q^n, \bar{Q}^n} (\Gamma_{n \delta_n} ({\calA}^{(n)}) \times {\calA}^{(n)}) +
P_{Q^n, \bar{Q}^n} (\Gamma_{n \delta_n} ({\calA}^{(n)}) \times {\calA}^{(n)c})
\\
\label{line1}
&&= P_{Q^n, \bar{Q}^n} (\Gamma_{n \delta_n} ({\calA}^{(n)}) \times {\calA}^{(n)})
\\
\label{line2}
&&=P_{Q^n, \bar{Q}^n} (d_H(Q^n, \bar{Q}^n) \leq n \delta_n )
\\
\nonumber
&&\geq 1-1/\lambda,
\end{eqnarray}
where ${\calA}^{(n)c}$ is the compliment set of $ {\calA}^{(n)}$, and  (\ref{line1}) and (\ref{line2}) follow from the fact that $P_{Q^n, \bar{Q}^n} (x^n, \bar{x}^n) =0$ when $\bar{x}^n \not \in {\calA}^{(n)}$. \qed

\subsection{Proof for Theorem \ref{theorem_blowingupwithConditionalEntropy}: Connection between Conditional Entropy and the Blowing-up Lemma}

We need two auxiliary lemmas. Recall that $\hat{Z^n}$ is the ``coloring" function on $Z^n$ at node $Z$.

\begin{lemma}
\label{lemma4}
Suppose $H(\hat{Z^n} | X^n)=n a_n$ with $a_n>0$. Then for all $\mu>1$,
$$Pr\left(X^n \in \left\{x^n: \, H(\hat{Z^n} |X^n=x^n) \leq n \mu a_n \right\} \right) \geq 1-1/\mu >0.$$
\end{lemma}

Proof: Define $A_n:=\{x^n: \, H(\hat{Z^n} |X^n=x^n) \leq n \mu a_n \}$. If $Pr(A_n)<1-1/\mu$, then we have
\begin{eqnarray*}
H(\hat{Z^n} |X^n ) &= &\sum_{x^n} Pr(x^n) H(\hat{Z^n} |X^n =x^n)
 \geq \sum_{x^n \in A_n^c} Pr(x^n) n \mu a_n
 >  \frac{1}{\mu}  n \mu a_n =n a_n.
\end{eqnarray*}
This is a contradiction. \qed

This lemma shows that for fixed $\mu>1$, there exist a constant $p_0>0$ and a set of codewords $A_n \in \calC^{(n)}$
such that: $Pr(X^n \in A_n)>p_0>0$, and $H(\hat{Z^n} |X^n=x^n) \leq \mu n a_n$ for all $x^n \in A_n$.

The following lemma characterizes the colors which have ``significant weight".
\begin{lemma} \label{lemma5}
Suppose $\{ p_j, j=1, \cdots, 2^{nR_0} \}$ is a probability distribution and $\sum_{j=1}^{2^{nR_0}} -p_j \log p_j \leq n b_n$. Then for
any $\alpha>1$, set $S:=\{j : \, p_j \geq 2^{-n b_n \alpha } \}$ has a total probability no less than $1-1/\alpha$.
\end{lemma}

Proof: We show that $S^c:=\{j : \, p_j < 2^{-n b_n \alpha } \}$ cannot have a total weight strictly larger than $1/\alpha$. We know
\begin{eqnarray*}
n b_n  & \geq  &\sum_j - p_j \log p_j
 \geq  \sum_{j \in S^c} - p_j \log p_j
\geq  \sum_{j \in S^c} -p_j \log(2^{-\alpha n b_n })
 =  \alpha n b_n   \cdot Pr(S^c).
\end{eqnarray*}
Thus the lemma is valid. \qed

We see that by making $\alpha$ to be a constant larger than one, the total weight of the colors with individual weight larger than $2^{-n a_n \alpha} $ is non-negligible.

We now present the proof for Theorem \ref{theorem_blowingupwithConditionalEntropy}.

{\bf Proof:}
By Lemma \ref{lemma4}, we know that for any $\mu>1$,
$$Pr \left(X^n \in  \{x^n: \, H(\hat{Z^n} |x^n) \leq n \mu a_n \} \right) \geq 1-1/\mu.$$

Define $\calC^{(n)}_1:=\{x^n: \, H(\hat{Z^n} |x^n) \leq n \mu a_n \}$. For each $x^n$ in $\calC^{(n)}_1$, by definition, we have
$\sum_{k=1}^{2^{nR_0}} -p_k \log p_k \leq n \mu a_n,$
where $p_k:= Pr(\hat{Z^n}=k |x^n)$.
Then by Lemma \ref{lemma5}, we know for any $\alpha>1$ there exists a set of colors $S$ such that: 1) $Pr( \hat{Z^n} \in S |x^n) \geq 1-1/\alpha$, and
2) For each color $j \in S$, $p_j \geq 2^{-n \mu a_n \alpha}$.

For such an $x^n$ and color $j$, by Lemma \ref{BlowupLemma}, the generalized blowing-lemma, we know for any $\lambda>1$,
$$Pr \left(\hat{Z^n} \in \Gamma_{n \lambda \sqrt{\mu a_n \alpha} }(\calA^{(n)}_j)  | X^n=x^n \right) \geq 1-1/\lambda.$$

Now the theorem is proved by letting $\mu=\alpha=\lambda$. \qed

\section{} \label{AppendixB}

\subsection{Proof for Theorem \ref{theorem_main0}}

By Theorem \ref{theorem_blowingupwithConditionalEntropy}, for any $\lambda>1$, there exist a set of code words $\calC^{(n)}_1$ and constant $p_0:=1-1/\lambda>0$ such that:
\begin{enumerate}
\item[i)]
 $Pr(X^n \in \calC^{(n)}_1) \geq p_0$; and
\item[ii)] For each $x^n\in \calC^{(n)}_1$, there exists a set of colors $S(x^n)$ such that
$Pr(\hat{Z^n} \in S(x^n) |x^n) \geq p_0$, and for each color $j \in S(x^n)$,
\begin{eqnarray}
\label{formula_a1}
Pr \left(Y^n \in \Gamma_{n \lambda^{3/2} \sqrt{a_n}} ( \calA^{(n)}_{j} ) |x^n \right)\geq p_0,
\end{eqnarray}
where $\calA^{(n)}_{j}:=\{z^n \in \Omega^n: \, \hat{z^n}=j\}$.
\end{enumerate}
Note we use $Y^n$ in (\ref{formula_a1}) instead of $Z^n$ because $Y^n$ and $Z^n$ are i.i.d. given $X^n$.
In other words, in the  ball of radius $n \lambda^{3/2} \sqrt{a_n}$ around an independently drawn $Y^n$, with probability at least $p_0$ one can find a point with color $j$, assuming the code word sent is from $\calC^{(n)}_1$.

Based on this, the following procedure can be applied to decode $X^n$ {\em solely based on $Y^n$}.
Randomly and uniformly pick a point $\omega^n$ in the ball centered at $Y^n$. Assume its color is  $\hat{\omega^n}$.
Apply the decoding function $f_n(\hat{\omega^n}, Y^n)$ to map to a code word, announce it the codeword decoded.

Now we calculate the decoding probability.
By assumption, since the code book is feasible, we have
$Pr(f_n(\hat{Z^n}, Y^n) \not= X^n) \rightarrow 0.$
So there exists an integer $N_1>0$ such that
\begin{eqnarray}
\label{formula_b1}
Pr(f_n(\hat{Z^n}, Y^n) \not= X^n ) \leq p_0^3/4, n>N_1.
\end{eqnarray}
We can also assume that at least half of the code words in $\calC^{(n)}_1$ satisfies
\begin{eqnarray}
\label{formula_b1_TOO}
Pr(f_n(\hat{Z^n}, Y^n) \not= X^n |X^n=x^n) \leq p_0, n>N_1.
\end{eqnarray}
Denote these code words as $\calC^{(n)}_2$.

For an $x^n \in \calC^{(n)}_2$ and a $j\in S(x^n)$, with $n\geq N_1$,
\begin{eqnarray*}
&&Pr \left(f_n(\hat{\omega^n}, Y^n) = X^n |x^n, \hat{Z^n}=j \right) \\
&& \geq  Pr \left( f_n(\hat{\omega^n}, Y^n)=X^n, \hat{\omega^n}=\hat{Z^n}, Y^n \in \Gamma_{n \lambda^{3/2} \sqrt{a_n}} (\calA^{(n)}_j) |x^n, \hat{Z^n}=j\right) \\
&& = Pr \left( f_n(\hat{Z^n}, Y^n)=X^n, \hat{\omega^n}=\hat{Z^n}, Y^n \in \Gamma_{n \lambda^{3/2} \sqrt{a_n}} (\calA^{(n)}_j) |x^n, \hat{Z^n}=j\right) \\
&& =  Pr \left( f_n(\hat{Z^n}, Y^n)=X^n, Y^n \in \Gamma_{n \lambda^{3/2} \sqrt{a_n}} (\calA^{(n)}_j) |x^n, \hat{Z^n}=j\right)  \\
&& \qquad \qquad \cdot
 Pr \left( \hat{\omega^n}=\hat{Z^n} \, | \, f_n(\hat{Z^n}, Y^n)=X^n, Y^n \in \Gamma_{n \lambda^{3/2} \sqrt{a_n}} (\calA^{(n)}_j), X^n=x^n, \hat{Z^n}=j\right) \\
&& \geq  Pr\left(f_n(\hat{Z^n}, Y^n)=X^n, Y^n \in \Gamma_{n \lambda^{3/2} \sqrt{a_n}} (\calA^{(n)}_j)|x^n, \hat{Z^n}=j \right)  \cdot \frac{1}{|Ball(n \lambda^{3/2} \sqrt{a_n})|},
\end{eqnarray*}
where the last inequality is because one picks a point $\omega^n$ uniformly within the ball which contains a point with color $\hat{Z}^n$.
We also know
\begin{eqnarray*}
&& Pr\left(f_n(\hat{Z^n}, Y^n)=X^n, Y^n \in \Gamma_{n \lambda^{3/2} \sqrt{a_n}} (\calA^{(n)}_j)|x^n,  \hat{Z^n}=j\right) \\
&&
=  Pr\left(Y^n \in \Gamma_{n \lambda^{3/2} \sqrt{a_n}} (\calA^{(n)}_j)|x^n, j \right) - Pr \left(f_n(\hat{Z^n}, Y^n) \not= X^n, Y^n \in \Gamma_{n \lambda^{3/2} \sqrt{a_n}} (\calA^{(n)}_j)|x^n, j \right)
\\
&&\geq
Pr(Y^n \in \Gamma_{n \lambda^{3/2} \sqrt{a_n}} (\calA^{(n)}_j)|x^n, j) - Pr(f_n(\hat{Z^n}, Y^n) \not= X^n|x^n, j)
\\
&& =
Pr(Y^n \in \Gamma_{n \lambda^{3/2} \sqrt{a_n}} (\calA^{(n)}_j)|x^n) - Pr(f_n(\hat{Z^n}, Y^n) \not= X^n|x^n, j),
\end{eqnarray*}
where the last equality comes from the fact that $Y^n$ is independent of $Z^n$ given $X^n$, and $\hat{Z^n}$ is a function of $Z^n$.
Because of (\ref{formula_a1}), we thus know
\begin{eqnarray}
\label{formula_a2}
&&Pr\left(f_n(\hat{\omega^n}, Y^n) = X^n |x^n, \hat{Z^n}=j \right)
\\
&&
\nonumber
\qquad  \qquad \qquad
\geq
\left(p_0 - Pr(f_n(\hat{Z^n}, Y^n) \not= X^n|x^n, j) \right)
  \cdot \frac{1}{|Ball(n \lambda^{3/2} \sqrt{a_n})|}.
\end{eqnarray}

Notice that (Recall $S(X^n)$ is the special color set of $X^n$)
$$Pr \left(X^n \in \calC^{(n)}_2, \hat{Z^n} \in S(X^n) \right) = Pr(X^n \in \calC^{(n)}_2) \cdot Pr \left(\hat{Z^n} \in S(X^n)| X^n \in \calC^{(n)}_2 \right) \geq p_0^2/2.$$
Thus, combining with (\ref{formula_a2}), we get
\begin{eqnarray*}
&&Pr(f_n(\hat{\omega^n}, Y^n) = X^n )
\\
&& = \sum_{x^n, j} Pr(X^n=x^n, \hat{Z^n}=j) \cdot Pr(f_n(\hat{\omega^n}, Y^n) = X^n | X^n=x^n, \hat{Z^n}=j )
\\
&& \geq \sum_{x^n \in \calC^{(n)}_2, j \in S(x^n)} Pr(X^n=x^n, \hat{Z^n}=j)
\cdot Pr(f_n(\hat{\omega^n}, Y^n) = X^n | X^n=x^n, \hat{Z^n}=j )
\\
&&
\geq    \frac{1}{|Ball(n \lambda^{3/2} \sqrt{a_n})|} \cdot \sum_{x^n \in \calC^{(n)}_2, j \in S(x^n)} Pr(X^n=x^n, \hat{Z^n}=j)
\cdot
\left( p_0 - Pr(f_n(\hat{Z^n}, Y^n) \not= X^n|x^n, j) \right)
\\
&&
= \frac{1}{|Ball(n \lambda^{3/2} \sqrt{a_n})|} \cdot
\\
&& \qquad \qquad \left(
Pr\left(X^n \in \calC^{(n)}_2, \hat{Z^n}\in S(X^n) \right) \cdot p_0 -
Pr\left(f_n(\hat{Z^n}, Y^n) \not= X^n, X^n \in \calC^{(n)}_2, \hat{Z^n}\in S(X^n) \right) \right)
\\
&& \geq
\frac{1}{|Ball(n \lambda^{3/2} \sqrt{a_n})|} \cdot \left(
p_0^3/2 -
Pr\left(f_n(\hat{Z^n}, Y^n) \not= X^n \right) \right)
\\
&&
\geq \frac{1}{|Ball(n \lambda^{3/2} \sqrt{a_n})|} \cdot p_0^3/4,
\end{eqnarray*}
where the last inequality is because of (\ref{formula_b1}).

By Arimoto's result (\ref{arimotoExponent}), we have
$\log|Ball(n \lambda^{3/2} \sqrt{a_n})|/n  +\delta_n \geq  \calE(R) ,$
where $\delta_n$ is a function of $n$ and $\lambda$.
\qed

\subsection{Derive $\calE(R)$ for the binary erasure channel}

For input distribution such that $Pr(X=0)=p$, by definition we have
\begin{eqnarray*}
&& \Phi_0(\rho, p)
\\
& =  &-\log \left[ \left( p (1-\epsilon)^{1/(1+\rho)} \right)^{(1+\rho)}
+ \left( (1-p) (1-\epsilon)^{1/(1+\rho)} \right)^{(1+\rho)} + \left( p \epsilon^{1/(1+\rho)} +(1-p) \epsilon^{1/(1+\rho)} \right)^{(1+\rho)} \right]\\
&=&
-\log \left[  p^{(1+\rho)} (1-\epsilon)
+  (1-p)^{(1+\rho)} (1-\epsilon) +  \epsilon \right] \\
&=&
-\log \left[  (p^{(1+\rho)} +  (1-p)^{(1+\rho)} ) (1-\epsilon) +  \epsilon \right].
\end{eqnarray*}

It is easy to show by checking the sign of $d\Phi_0(\rho, p) /dp$ that
$\mbox{argmin}_p \Phi_0(\rho, p) = 1/2, $ noting $\rho<0$.
And thus
\begin{eqnarray*}
\calE(R) &= &\max_{\rho \in [-1, \,\, 0)} \left(-\rho R + \min_p \Phi_0(\rho, p)\right)
 = \max_{\rho \in [-1, \,\, 0)} \left(-\rho R  -\log [2^{-\rho} (1-\epsilon) + \epsilon]\right)
\\
&
= & \max_{x \in (0, 1]}  \left( Rx  -\log [2^x (1-\epsilon) + \epsilon] \right)
\\
&=: &\max_{x \in (0, 1]} s(x).
\end{eqnarray*}

The derivative of $s(x)$ is
$R  - \frac{2^x (1-\epsilon)}{2^x (1-\epsilon) + \epsilon}.$
We observe that: 1) when $x=0$, $s'(x)=R-(1-\epsilon)>0$, as $R$ is larger than $C_{XY}$; and
2) $s'(x)$ is monotonically decreasing in $x$.
For $s'(x)$ to be zero, one must have
$2^x=\frac{R \epsilon}{(1-\epsilon)(1-R)}.$
Hence we know that for $R \in (1-\epsilon, 1-\frac{\epsilon}{2-\epsilon})$, $s'(x)$ can reach 0. When $R>1-\frac{\epsilon}{2-\epsilon}$, $s'(x)$ is always positive.
In the latter case, we know $\calE(R) = R- \log (2-\epsilon).$

In sum, we know
$$\calE(R) = \left\{
\begin{array}{l}
R \log \frac{R \epsilon}{(1-\epsilon) (1-R)} -\log \left(\frac{R \epsilon}{1-R} +\epsilon \right),
 \,\, R \in  (1-\epsilon, 1-\frac{\epsilon}{2-\epsilon})\\
R- \log (2-\epsilon), \,\, R \geq 1-\frac{\epsilon}{2-\epsilon}.
\end{array}
\right.$$

\section{Proof for Theorem  \ref{theorem_simuSideInfo}}
\label{AppendixC}

We first need the following, which is almost the same as known result in \cite{Cuff2008}.

\begin{lemma} \label{lemma_chanSimuSpecificYn}
Suppose $Y^n$ is a constant for each $n$, i.e., $Y^n \equiv y_0^n$. Then for any $\delta>0$, there exist simulation encoding and decoding with rate $R_1=\limsup i(X^n; Z^n | y_0^n) / n + \delta = \limsup \frac{1}{n} \log \frac{\bar{p}(Z^n|X^n, y_0^n)}{\bar{p}(Z^n | y_0^n)}+\delta$ and $R_2$ sufficiently large such that
$$d(X^n, Z^n | Y^n=y_0^n):= \sum_{x^n, z^n} \left| \bar{p}(x^n, z^n | y_0^n) - \bar{p}(x^n | y_0^n) Q(z^n|x^n, y_0^n) \right| \to 0.$$
\end{lemma}

Proof: The theorem is the same as a special case in the achievability part in Section VI of \cite{Cuff2008}. As stated in Remark \ref{remark_diff2CuffFormulation},
there are two small differences between our formulation and that in \cite{Cuff2008}. First,  the source $X^n$ is not generated from i.i.d. random variables $X_1, X_2, \cdots, X_n$ based on a distribution $p(x)$. Instead, here the source is uniformly picked from an existing code book. Secondly, there is a side information $Y^n$ in our formulation.

Replacing the Lemma 6.1 in \cite{Cuff2008} with (our) Lemma \ref{Lemma_HanVerdu}  (which is copied from \cite{HanVerdu1993} ) and making $R_2$ large, the same proof goes through\footnote{Actually Lemma 6.1 in \cite{Cuff2008} is a simplified version of Lemma \ref{Lemma_HanVerdu}.}. Note that the mutual information becomes the corresponding limsup expression. As a remark, for a given $y_0^n$, the channel is a ``conditioned" channel. The source distribution and channel distribution (for given input) may be different than without conditioning, but Lemma \ref{Lemma_HanVerdu} still applies. This is because everything is conditioned on $y_0^n$.
\qed

Now we are ready to prove Theorem  \ref{theorem_simuSideInfo}.

{\em Proof for Theorem \ref{theorem_simuSideInfo}:}
We know that
\begin{eqnarray*}
&& \sum_{x^n, y^n, z^n} \left| \bar{p}(x^n, y^n, z^n) - \bar{p}(x^n, y^n) Q(z^n|x^n, y^n) \right|  \\
&& = \sum_{y^n} \bar{p}(y^n) \sum_{x^n, z^n} \left| \bar{p}(x^n, z^n |y^n) - \bar{p}(x^n | y^n) Q(z^n|x^n, y^n) \right| \\
&& =  \sum_{y^n} \bar{p}(y^n) \cdot  d(X^n; Z^n | Y^n=y^n).
\end{eqnarray*}
One can apply the channel simulation procedure for each $y^n$ as indicated in Lemma \ref{lemma_chanSimuSpecificYn}. (The detailed simulation procedure is in Section VI of \cite{Cuff2008}.) This leads to a simulated distribution $\bar{p}(x^n, y^n) Q(z^n |x^n, y^n)$.

If
$ \sum_{y^n} \bar{p}(y^n) \sum_{x^n, z^n} \left| \bar{p}(x^n, z^n |y^n) - \bar{p}(x^n | y^n) Q(z^n|x^n, y^n) \right| \nrightarrow  0, \, n \to \infty,$
then there exist a series of integers $n_k$ going to infinity, positive constants $c_1$ and $c_2$, and events $A_{n_k} \subseteq \Omega_Y^{n_k}$ such that:
\begin{enumerate}
\item $Pr(Y^{n_k} \in A_{n_k})>c_1>0$;
\item for all $y^{n_k} \in A_{n_k}$, $d(X^{n_k}; Z^{n_k} | Y^{n_k}=y^{n_k})>c_2$.
\end{enumerate}
By Lemma \ref{lemma_chanSimuSpecificYn}, we must have
\begin{eqnarray}
\label{forProofTheorem9}
\limsup i(X^{n_k}; Z^{n_k} | Y^{n_k}=y^{n_k})/n \geq R_1, \forall y^{n_k} \in A_{n_k}.
\end{eqnarray}

On the other hand, since $R_1 = \limsup i (X^n; Z^n|Y^n)/n +\delta$, by definition we have
\begin{eqnarray*}
&& Pr\left(\frac{1}{n} \log \frac{ \bar{p} ( X^n, Z^n |Y^n) }{ \bar{p}(X^n |Y^n) } \leq R_1-\delta/2 \right)  \to 1, n \to \infty.
\end{eqnarray*}
Because
\begin{eqnarray*}
&& Pr\left(\frac{1}{n} \log \frac{\bar{p} ( X^n, Z^n |Y^n)}{ \bar{p}(X^n |Y^n)} \leq R_1- \frac{\delta}{2} \right) =
\sum_{y^n} \bar{p}(y^n) Pr\left(\frac{1}{n} \log \frac{\bar{p}(X^n, Z^n |y^n) }{ \bar{p}(X^n |y^n) } \leq R_1-\frac{\delta}{2} \, | \, Y^n=y^n \right),
\end{eqnarray*}
we know that the set of $y^n$'s such that $\limsup i(X^{n}; Z^{n} | Y^{n}=y^{n})/n \geq R_1$ must have a probability going to zero, as $n$ goes to infinity.
This is a contradiction to (\ref{forProofTheorem9}) above! \qed

\section{}
\label{AppendixD}

\subsection{Proof for Lemma \ref{lemma_i2Irelation}}

By definition, $E i(X^n; Y^n)=I(X^n; Y^n)$. Since $\liminf \frac{1}{n} i(X^n; Y^n) \geq \bar{C}_{XY}-c_1$ , for any $\delta>0$, there exists a sequence $\{n_k\}$ and $p_\delta>0$ such that
\begin{eqnarray}
\label{equation_appendixD1}
Pr(i(X^{n_k}; Y^{n_k})/n_k \leq \bar{C}_{XY} -c_1 +\delta ) \geq p_\delta.
\end{eqnarray}

Furthermore, for any $n$ we have
\begin{eqnarray}
\nonumber
i(X^n; Y^n)/n &= &\frac{1}{n} \log \frac{p(Y^n|X^n)}{p(Y^n)}
\\
\nonumber
&= &\frac{1}{n} \log \frac{p(Y^n|X^n)}{\sum_{x^n} p(x^n) p(Y^n | x^n)} \\
\nonumber
&= &\frac{1}{n} \log \frac{p(Y^n|X^n)}{2^{-nR} \sum_{x^n} p(Y^n | x^n)} \\
\label{equation_appendixD2}
& \leq & \frac{1}{n} \log 2^{nR} = R.
\end{eqnarray}

Also, since the channel $XY$ satisfies the strong converse property, by Lemma 10 of \cite{HanVerdu1993}, we know
there exists $\epsilon_k \to 0$ such that
\begin{eqnarray}
\label{equation_appendixD3}
Pr\left( \frac{i(X^n; Y^n)}{n} \leq \bar{C}_{XY} +\epsilon_k \right) \to 1, \quad n \to \infty.
\end{eqnarray}

Combining the above (\ref{equation_appendixD1}), (\ref{equation_appendixD2}), (\ref{equation_appendixD3}), we know
\begin{eqnarray*}
I(X^{n_k}; Y^{n_k})/{n_k} &=&E i(X^{n_k}; Y^{n_k}) /{n_k} \\
& \leq  & (\bar{C}_{XY} -c_1 +\delta) p_\delta + (\bar{C}_{XY}+\epsilon_k) (1-p_\delta - o(1)) + R \cdot o(1) \\
& =&
\bar{C}_{XY} -(c_1 - \delta) p_\delta + o(1).
\end{eqnarray*}
The lemma is proved by defining $c_2=\max_{\delta \geq 0} (c_1 - \delta) p_\delta$.\qed

\subsection{Proof for Theorem \ref{theorem_mainGeneral} }

First focus on the relay channel. Denote $a_n:=H(\hat{Z^n}|X^n) / n$.
For any $n$, we know $H(X^n)=nR$ and, by Fano's lemma, $H(X^n|Y^n, \hat{Z^n})= n \cdot o(1)$. We have
\begin{eqnarray}
\nonumber
&& n(R+o(1))=I(X^n; Y^n, \hat{Z^n}) \\
&&
\nonumber
=I(X^n; Y^n) +H(\hat{Z^n}|Y^n)-H(\hat{Z^n}|X^n) \leq I(X^n; Y^n) +nR_0 -n a_n.
\end{eqnarray}
That is, $R\leq I(X^n; Y^n)/n +R_0 - a_n +o(1)$.
By Lemma \ref{lemma_i2Irelation}, we know there exits $n_k$ going to infinity such that
\begin{eqnarray}
\label{def_ank}
R\leq C_{XY}-c_2 +R_0 - a_{n_k} +o(1).
\end{eqnarray}
Denoting $a:=\limsup a_{n_k}$, we have
\begin{eqnarray}
\label{Theorem_equation1}
R\leq C_{XY}-c_2 +R_0 - a.
\end{eqnarray}
This satisfies i) and iii).

Now we show ii) by applying results from channel simulation for the companion channel $\bar{p}(y, z|x)$.

One can simulate the companion channel $\bar{p}(y, z|x)$ as follows. By Lemma \ref{lemma_fromixy2ixzgivenY} and Theorem \ref{theorem_simuSideInfo}, for any $\delta>0$, with rate
\begin{eqnarray}
\label{def_R1}
R_1:= \bar{C}_{XYZ}-C_{XY}+c_1+\delta,
\end{eqnarray}
 one can encode the channel $\bar{p}(x^n, y^n, z^n)$ based on side information $Y^n$ and common randomness $K$, by applying $\phi_n(X^n, Y^n, K)$.
Given $U$, the channel decoder generates an output $\tilde{Z}^n$ based on $U$, $Y^n$ and $K$ by function $\psi_n(U, Y^n, K)$. And we know that
$$\sum_{x^n, y^n, z^n} | \bar{p}(x^n, y^n, z^n) - \bar{p}(x^n, y^n) Q(z^n|x^n, y^n)| \to 0, n \to \infty.$$

In the relay channel, node $Y$ can utilize this channel simulation to produce a $\tilde{Z^n}$ with distribution close to that of the relay's observation $Z^n$ as follows. To generate a channel simulation output it needs $U, K, Y^n$. It has $Y^n$ because it observes it directly. It has $K$ because it is a common randomness -- a random variable uniformly distributed on $\{1, 2, \cdots, 2^{nR_2}\}$. For $U$, there are total $2^{n R_1}$ possibilities. Node $Y$ picks a $\tilde{U}$ uniformly in $\{1, 2, \cdots, 2^{nR_1} \}$, and generates a $\tilde{Z}^n$ based on $\psi_n(\tilde{U}, Y^n, K)$. Note that the probability to hit the correct $U$ is $2^{-n R_1}$.

Assume that $\tilde{Z}^n$ is the channel simulation output.
Node $Y$ can apply the same procedure and argument as in Section IV to guess $X^n$. Specifically, it draws a ball of radius $n \lambda^{3/2} \sqrt{a_n}$ around $\tilde{Z^n}$ in the space $\Omega_Z^n$, for a constant
$\lambda>1$. Then it picks a point $\omega^n$ uniformly in the ball and applies the known decoding function $f_n(\hat{\omega}^n, Y^n)$ to guess $X^n$.

Now we analyze the decoding probability. Suppose $\tilde{Z^n_1}$ is a random variable such that it is i.i.d. of $Z^n$ conditioned on $X^n$. We know as in Section IV that, when $X^n$ is from a non-diminishing set of code words  $\calC^{(n)}_1$, the ball around $\tilde{Z}^n_1$ of radius $n \lambda^{3/2} \sqrt{a_n}$ will contain a point with the same color as $Z^n$'s with provability no less than a positive constant $p_0$. To be specific, assume $\hat{Z}^n=\hat{c}$ and  define
$$\Gamma_{\hat{c}}^{n \lambda^{3/2} \sqrt{ a_n}}:= \left\{ y^n \in \Omega_Y^n: \, \mbox{ There is $y_1^n$ with color $\hat{c}$ and $d_H(y^n, y_1^n) \leq n \lambda^{3/2} \sqrt{ a_n}$ } \right\}.$$
That is, $\Gamma_{\hat{c}}^{n \lambda^{3/2} \sqrt{ a_n}}$ is the blown-up set of the points with color $\hat{c}$.

We have
$$Pr \left( \tilde{Z^n_1} \in \Gamma_{\hat{c}}^{n \lambda^{3/2} \sqrt{ a_n}} | X^n \in \calC^{(n)}_1 \right) \geq p_0.$$

However the above analysis is for a hypothetical $\tilde{Z}_1^n$. What the channel simulation really generates is $\tilde{Z}^n$. For $\tilde{Z}^n$, because of (\ref{formula_theorem10}), we know
$$Pr \left(\tilde{Z^n} \in \Gamma_{\hat{c}}^{n \lambda^{3/2} \sqrt{ a_n}} | X^n \in \calC^{(n)}_1 \right) = Pr \left(\tilde{Z^n_1} \in \Gamma_{\hat{c}}^{n \lambda^{3/2} \sqrt{ a_n}} | X^n \in \calC^{(n)}_1 \right)+o(1) \geq p_0+o(1).$$
That is, the ball around  $\tilde{Z}^n$ of radius $n \lambda^{3/2} \sqrt{a_n}$ still contains the color of $Z^n$ with non-diminishing probability.

Now we can bound the decoding probability as follows, as in Section IV.
$$Pr(\mbox{Node $Y$ can decode correctly}) \geq  2^{-n R_1} \cdot (\mu_1+o(1)) \frac{1}{|Ball(n \lambda^{3/2} \sqrt{ a_n})|},$$
where $\mu_1>0$ is a function of $p_0$.
In words, it is saying that one can decodes if the correct $U$ is used and the correct color is hit in the ball.
Based on the result of Arimoto \cite{Arimoto1973}, one must have
$$\calE_Y(R) \leq R_1 +\limsup \frac{1}{n_k} \log \left| Ball(n \lambda^{3/2} \sqrt{ a_{n_k}}) \right|.$$
Recall $n_k$ and $R_1$ are defined in (\ref{def_ank}) and (\ref{def_R1}), respectively. Letting $\lambda$ go to one and then $k$ go to infinity, we get

$$\calE_Y(R) \leq \bar{C}_{XYZ}-C_{XY}+c_1 +H_2(\sqrt{a}) + \sqrt{a} \log |\Omega_Z|.$$
\qed


\begin{thebibliography}{99}

\bibitem{Meulen1971} E. C. Van der Meulen, ``Three-terminal communication channel," {\it Adr. Appl. Proh.}. vol. 3, pp. 120-154, 1971.

\bibitem{ElGamal2010isit} A. El Gamal, ``Coding for noisy networks," plenary talk at the 2010 IEEE International Symposium on Information Theory, Austin, Texas, USA

\bibitem{CoverElGamal1979} T.M. Cover and A. El Gamal, ``Capacity theorems for the relay channel," {\it IEEE Trans. Inform. Theory}, vol. 25, no. 5, pp. 572-584, Sept. 1979.

\bibitem{CoverThomas1991} T.M. Cover and J.S. Thomas, Elements of Information Theory. New
York: Wiley, 1991.

\bibitem{ElGamalAref1982} A. El Gamal and M. Aref, ``The capacity of the semideterministic relay channel," {\it IEEE Trans. Inf. Theory}, vol. IT-28, no. 3, p. 536, May 1982.

\bibitem{Kim2008} Y. H. Kim, ``Capacity of a class of deterministic relay channels," in {\it IEEE Trans. on Information Theory}, 2008.



\bibitem{AlswedeGacsKorner1976} R. Ahlswede, P. Gacs, and J. Korner, ``Bounds on conditional probabilities
with application in multi-user communication,", {\it Z. Wuhrschernlrchkertstheorie
uenv. Gehiete}, vol. 34, pp. 157-177, 1976.

\bibitem{AleksicRazaghiYu2009} M. Aleksic, P. Razaghi, and W. Yu, ``Capacity of a class of modulo-sum relay channels," {\it IEEE Trans. Inf. Theory}, vol. 55,
no. 3, pp. 921-930, 2009.


\bibitem{Arimoto1973}
S. Arimoto, ``On the converse to the coding theorem for discrete memoryless channels," {\it IEEE Trans. Info. Theory},  Vol. 19, No. 3, pp.357-359, May 1973

\bibitem{Marton1986} K. Marton, ``A simple proof of the Blowing-Up Lemma," {\it IEEE Trans. on Information Theory,} IT-32, pp. 445-446, 1986.

\bibitem{Omstein} D. Omstein, {\it Ergodic Theory, Randomness and Dynamical Systems}, Yale Univ. Press: New Haven, 1974.

\bibitem{Zhang1988} Z. Zhang, ``Partial converse for a relay channel," {\it IEEE Trans. Info. Theory,} Vol. 34, No. 5, pp. 1106-1110, September 1988.

\bibitem{Wyner1975} A. Wyner, ``The Common Information of Two Dependent Random
Variables," in {\it IEEE Trans. Info. Theory}, vol. IT-21, no. 2, March 1975.

\bibitem{HanVerdu1993} T.S. Han and S. Verdu, ``Approximation theory of output statistics, " {\it IEEE Trans. Info. Theory, } Vol. 39, No. 3, pp. 752-772, May 1993.

\bibitem{Cuff2008} P. Cuff, ``Communication requirements for generating correlated random variables," {\it ISIT 2008}.





\bibitem{ElGamal2010} A. El Gamal, Talk on Katalin Marton's work at the third WITHITS Annual Event at ISIT 2010.
http://isl.stanford.edu/~abbas/presentations/Marton.pdf

\end{thebibliography}
\end{document}